%% file: ms.tex
\documentclass[sigconf,nonacm]{acmart}

\usepackage{graphicx}
\usepackage{subfiles}

\usepackage{todonotes}
\usepackage{booktabs}
\PassOptionsToPackage{hyphens}{url}\usepackage{hyperref}

\usepackage{dsfont}
\usepackage{amsmath}
\usepackage{nicefrac}

\usepackage{caption}
\usepackage{subcaption}
\usepackage{tablefootnote}
\usepackage{pifont}
\newcommand{\cmark}{\textcolor{teal}{\ding{51}}}%
\newcommand{\xmark}{\textcolor{red}{\ding{55}}}%

\usepackage{multicol}
\usepackage{multirow}
\usepackage{url}

\usepackage{tikz}
\tikzset{
  treenode/.style = {shape=rectangle, rounded corners,
                     draw, align=center,
                     top color=white, bottom color=white},
  root/.style     = {treenode, font=\Large, bottom color=red!30},
  env/.style      = {treenode, font=\normalsize},
  dummy/.style    = {circle,draw}
}

\usepackage{xcolor}

\newcommand\ml[1]{\textcolor{teal}{#1}}

\AtBeginDocument{%
  \providecommand\BibTeX{{%
    \normalfont B\kern-0.5em{\scshape i\kern-0.25em b}\kern-0.8em\TeX}}}

\begin{document}

\title{Overview of the TREC 2022 NeuCLIR Track}

\author{Dawn Lawrie$^\dagger$, Sean MacAvaney$^\ddagger$, James Mayfield$^\dagger$, \\ 
Paul McNamee$^\dagger$, Douglas W. Oard${^o}^\dagger$, Luca Soldaini$^*$, Eugene Yang$^\dagger$}
\affiliation{
  \institution{$^\dagger$John Hopkins University Human Language Technology Center of Excellence,\\
  $^\ddagger$University of Glasgow, $^o$University of Maryland, $^*$Allen Institute for AI}
  \country{}
}

\email{lawrie@jhu.edu,sean.macavaney@glasgow.ac.uk, james.mayfield@jhuapl.edu}
\email{mcnamee@jhu.edu, lucas@allenai.org, oard@umd.edu, eugene.yang@jhu.edu}

\renewcommand{\shortauthors}{Lawrie et al.}

\begin{abstract}
This is the first year of the TREC Neural CLIR (NeuCLIR) track,
which aims to study the impact of neural approaches to cross-language information retrieval.
The main task in this year's track was ad hoc ranked retrieval of Chinese, Persian, or Russian newswire documents
using queries expressed in English.
Topics were developed using standard TREC processes,
except that topics developed by an annotator for one language were assessed by a different annotator
when evaluating that topic on a different language. 
There were 172 total runs submitted by twelve teams.
\end{abstract}

\settopmatter{printfolios=true}
\maketitle

\input{1-intro.tex}

\input{2-task.tex}
\input{3-data.tex}

\input{4-topic.tex}

\input{5-judgments.tex}

\input{6-participation.tex}

\input{7-results.tex}

\input{8-future.tex}

\input{9-conclusion.tex}

\bibliographystyle{ACM-Reference-Format}
\bibliography{sigirforum}

\input{_tab-full-results-zho.tex}
\input{_tab-full-results-fas.tex}
\input{_tab-full-results-rus.tex}

\input{_fig_topic_box}

\input{_fig_overlaps.tex}

\end{document}

%% file: 1-intro.tex
\section{Introduction}

Cross-language Information Retrieval (CLIR) has been studied for more than three decades, first appearing at the Text Retrieval Conference (TREC) in TREC-4~\cite{davis1995trec}.
Prior to the application of deep learning,
strong statistical approaches were developed that work well across many languages.
Though as with most other language technologies,
neural methods have led to substantial improvements in information retrieval.
Several factors combined to make us feel that the time was right to press for rapid progress in CLIR:

\begin{itemize}
\itemsep0em 

    \item \textbf{Research Community.} There have been recent programs focused on CLIR such as IARPA MATERIAL\footnote{\url{https://www.iarpa.gov/research-programs/material}} and the Johns Hopkins Human Language Technology Center of Excellence (HLTCOE) Summer Camp for Applied Language Engineering (SCALE) 2021\footnote{\url{https://hltcoe.org/research/scale}}.  Recent interest among natural language processing researchers in the related problems of cross-language question answering and development of multilingual embeddings 
    have produced a new crop of researchers familiar with and interested in CLIR and related tasks.

    \item \textbf{Algorithms.} Neural advances in the state of the art in monolingual retrieval have been appearing for several years.
    Improvements in cross-language IR have come just in the last year or two.
 
    \item \textbf{Data.} The appearance of MS MARCO led to rapid advances in monolingual IR.
    Translations of MS MARCO into other languages have allowed training of CLIR systems.
    Additional resources that could also be useful for training neural CLIR models have also appeared recently, including
    CLIRMatrix~\cite{clirmatrix},\footnote{\url{https://github.com/ssun32/CLIRMatrix}}
    HC4~\cite{hc4},\footnote{\url{https://github.com/hltcoe/hc4}}
    WikiCLIR~\cite{wikiclir},\footnote{\url{https://www.cl.uni-heidelberg.de/statnlpgroup/wikiclir/}}
    and MIRACL~\cite{miracl}.\footnote{\url{https://github.com/project-miracl/miracl}}

    \item \textbf{Infrastructure.} Earlier systems for experimental IR have recently been supplemented by systems such as PyTerrier~\cite{pyterrier}\footnote{\url{https://github.com/terrier-org/pyterrier}} and Castorini\footnote{\url{https://github.com/castorini/}} that support neural methods, and by systems such as Patapsco~\cite{costello22-patapsco}\footnote{\url{https://github.com/hltcoe/patapsco}} that are designed specifically for CLIR.
    These systems provide a base on which to build, somewhat lowering barriers to entry, and providing a source for baselines to which progress can be compared.

\end{itemize}
The NeuCLIR track was designed to take advantage of this confluence of interest and resources
to push the state of the art in neural CLIR forward.
We expect the track to help to answer at least the following questions:
\begin{itemize}
\itemsep0em 
    \item What are the best neural CLIR approaches?
    \item How do the best approaches compare to the straightforward combination of machine translation and monolingual IR?
    \item How do the best neural approaches compare to the strongest statistical approaches to CLIR?
    \item Can reranking further improve retrieval effectiveness using techniques that would be impractical for full-collection retrieval?
    \item How do the resource requirements for the various approaches compare?
    \item What resources are most useful for training CLIR systems?
    \item What are the best neural multilingual information retrieval (MLIR) approaches for producing a single ranked lists containing documents in several languages?
\end{itemize}
NeuCLIR 2022 has helped start to answer these questions.
The track will continue in 2023.

The NeuCLIR track maintains an official website at: \url{https://neuclir.github.io}.

%% file: 2-task.tex
\section{Task Definition}

We explore three tasks in the TREC 2022 NeuCLIR track:
ad hoc CLIR, reranking CLIR, and monolingual.
All three tasks use the same document collections, topics, and relevance assessments.
Monolingual runs use topics manually translated into the language of the documents;
ad hoc and reranking runs use the original English topics.
Ad hoc runs rank documents from the entire collection,
while reranking runs rank only the 1,000 documents that appear in the output of a NIST-provided initial run.

\subsection{Ad Hoc CLIR Task}
The main task in the NeuCLIR track is ad hoc CLIR
Systems receive a document collection in Chinese, Persian, or Russian,
and a set of topics written in English.
For each topic, the system must return a ranked list of 1,000 documents drawn from the entire document collection of the target language,
ordered by likelihood and degree of relevance to the topic.
Runs that use a human in the loop for ad hoc retrieval
(or had design decisions influenced by human review of the topics)
are indicated as ``manual'' runs;
all others are considered ``automatic.''

\subsection{Reranking CLIR Task}
The reranking task provides an initial ranked list of 1,000 retrieved documents from the document collection.
Each ranked list is the output of a BM25 retrieval system, which used document translation to cross the language barrier. The run ids are {\texttt{coe22-bm25-td-dt-*}}  where {\texttt{*}} is {\texttt{zho}} for Chinese, {\texttt{fas}} for Persian, and {\texttt{rus}} for Russian. The runs appear in bold in Tables~\ref{tab:zho-full-results}, \ref{tab:fas-full-results}, and \ref{tab:rus-full-results}. These runs use the English title and descriptions for queries. 
Systems are then asked to rerank the documents to produce a new ordering that improves an evaluation metric.
This task is suitable for teams that want to focus on second-stage scoring models,
rather than on models which search an entire collection.

\subsection{Monolingual Retrieval Task}
While monolingual retrieval is not a focus of the NeuCLIR track,
monolingual runs can improve assessment pools
and serve as good points of reference for cross-language runs.
The monolingual retrieval task is identical to the ad hoc task,
but it uses topic files that are human translations of the English topics
into a target language in a way that would be expressed by native speakers of the language.
This task is suitable for teams looking to explore monolingual ranking in languages other than English.
It is also a lower barrier to entry task for teams that are interested in the track.

%% file: 3-data.tex
\section{Documents}

There are three components of the 2022 NeuCLIR test collection: documents, topics, and relevance judgments.  In this section we describe the documents. 

The document collection, NeuCLIR-1,  consists of documents in three languages: Chinese, Persian, and Russian,
drawn from the Common Crawl news collection.\footnote{\url{https://commoncrawl.org/2016/10/news-dataset-available/}}
The documents were obtained by the Common Crawl service between August 1, 2016 and July 31, 2021;
most of the documents were published within this five year window.
Text was extracted from each source web page using the Python utility \textit{Newspaper}.\footnote{\url{https://github.com/codelucas/newspaper}}

While NIST made the documents (and topics) available for participants and will distribute them for the
foreseeable future, an alternative source of the document collection can be obtained 
directly from Common Crawl, which is
the original source.
A github repository\footnote{\url{https://github.com/NeuCLIR/download-collection}} facilitates this
by providing code to access the documents via their Universally Unique Identifiers (UUID).
The process extracts the text from the documents and then matches the descriptor
to ensure that both track participants and non-participants index the same documents. 

The collection was distributed to participants in JSONL,
a list of JSON objects, one per line.
Each line represents a document.
Each document JSON structure consists of the following fields:
\begin{description}
\itemsep0em 
    \item [id:] UUID assigned by Common Crawl %
    \item [cc\_file:] raw Common Crawl document
    \item[time:] time of publication, or null
    \item[title:] article headline or title
    \item[text:] article body
    \item[url:] address of the source web page
\end{description}

To ascertain the language of each document,
its title and text were independently run through two automatic language identification tools, \textit{cld3}\footnote{\url{https://pypi.org/project/pycld3/}} and an in-house tool, \textit{VALID} \cite{mcnamee-2016-language}, a compression-based model trained on Wikipedia text.
Documents for which the tools agreed on the language,
or where one of the tools agreed with the language recorded in the web page metadata
were included in the collection under the language of agreement;
all others were removed. 
This is an imperfect process and some documents comprised of text in other languages are in the collections.
The extent of the language pollution is unknown;
however, annotators did sometimes encounter out-of-language documents in the pools of assessed documents.
These documents were always considered not relevant.
While we expected this process to make errors,
we had assumed that no systems would retrieve out-of-language documents.
This assumption proved to be false as some systems ranked documents in other languages highly.
All documents with more than 24,000 characters (approximately 10 pages of text) were also removed, as very long documents create challenges in assessment.
Additionally, very short documents were also removed, specifically: Chinese documents containing 75 or fewer characters, Persian documents containing 100 or fewer characters, and Russian documents containing 200 or fewer characters.
We observed that such documents are often not genuine news articles, frequently consisting of isolated headlines or commercial advertisements.

Each collection was limited in size to at most 5 million documents.
After removing duplicates, the Russian collection was significantly above this threshold.
Therefore, we used Scikit-learn's implementation of random sampling without replacement\footnote{\url{https://scikit-learn.org/stable/modules/generated/sklearn.utils.random.sample_without_replacement.html}} to downsample the collection.
Final collection statistics appear in Table~\ref{tab:coll_stats}.

\begin{table*}[tb]
\centering
\caption{Document Collection Statistics for NeuCLIR-1 (token counts from Spacy)}.\label{tab:coll_stats}
\begin{tabular}{cccccc}
\toprule
 & \multicolumn{1}{c}{Document}  & \multicolumn{1}{c}{Avg. Chars}  & \multicolumn{1}{c}{Median Chars}  & \multicolumn{1}{c}{Avg. Tokens}  & \multicolumn{1}{c}{Median Tokens}  \\ 
Language &  \multicolumn{1}{c}{count} & \multicolumn{1}{c}{per Document} & \multicolumn{1}{c}{per Document} &  \multicolumn{1}{c}{per Document} & \multicolumn{1}{c}{per Document} \\ 
\midrule
Chinese & 3,179,209 & 743 & 613 & 427 & 356 \\
Persian & 2,232,016 & 2032 & 1427 & 429 & 300 \\
Russian & 4,627,543 & 1757 & 1198 & 301 & 204 \\ 
\bottomrule
\end{tabular}
\end{table*}

%% file: 4-topic.tex
\section{Topics}

NeuCLIR 2022 topics were developed to be traditional TREC-style information needs that are broader than CLIR question answering, which can be answered with a 
phrase or sentence.
Topic development was done in two phases. %
First, assessors created a topic by writing an English description
and searching for relevant documents in their language.
Subsequently, pools were created from track submissions for assessors to judge.
This section focuses on topic development;
Section~\ref{sec:rel-judge} addresses relevance assessment.

During topic development, assessors wrote the title, description, and narrative components of the topic
and then examined thirty documents that they discovered through interactive monolingual search
using a web interface to the Patapsco framework~\cite{costello22-patapsco}.
They recorded the number of relevant documents they discovered.
Any topic where the assessor recorded more than twenty relevant documents 
was deemed too productive for inclusion in the collection;
such topics are deleterious to collection reusability because they can lead to too many relevant documents being unjudged.
With the idea of adding a multilingual task to NeuCLIR 2023,
we wanted topics that would exhibit relevant documents in multiple languages. 
Assessors therefore also judged documents for topics developed by assessors in other languages.

This process developed 137 topics.
A total of 89 of these were evaluated in Chinese, 69 in Persian, and 104 in Russian.
All topics were examined by one of the organizers;
topics that were overly ambiguous, too similar to another topic, or judged by the organizers to be otherwise inappropriate were eliminated.
A total of 114 topics remained; these were distributed to participants.
Human translations of each topic into the three collection languages were produced,
and each was subsequently vetted by a language expert to ensure translation quality.
These translations were used for the monolingual task.

Given that no language assessed all 137 topics, there was 
insufficient information available after the first phase of topic development 
by NIST to select multilingual topics. 
To further evaluate topics and in particular to identify topics that would have some but not too many relevant documents, additional assessors were used
to judge the topics.
These additional assessors, who will be referred to as {\it{non-NIST assessors}},
had proficiency in the language for which they provided annotation,
but generally were not native speakers of the language.
The non-NIST assessors judged 63 topics in Chinese, 67 in Persian, and 68 in Russian.
They used the Patapsco framework to perform monolingual retrieval using BM25;
however, the interface for interacting with the system was different from that used by the NIST assessors.
Two main differences were that non-NIST assessors provided document judgments with the interface,
rather than reporting a count;
and they had access to HiCAL~\cite{hical},
an active learning system, to recommend documents for assessment.
While non-NIST assessors were encouraged to judge thirty documents,
the actual number varied between five and sixty-two,
with a median of fifteen documents judged per topic.
Assessors identified between zero and eighteen relevant documents with an average of 4.6 documents per topic.
Because of the lack of consistency
of the number of judged documents per topic, rather than using a cutoff of 20 documents
out of 30 as a sign of too many relevant documents, 
any topic with more than 65\% relevant documents was deemed too productive to contribute to a reusable collection. 

\begin{figure*}[tb]
\centering
\includegraphics[width=\linewidth]{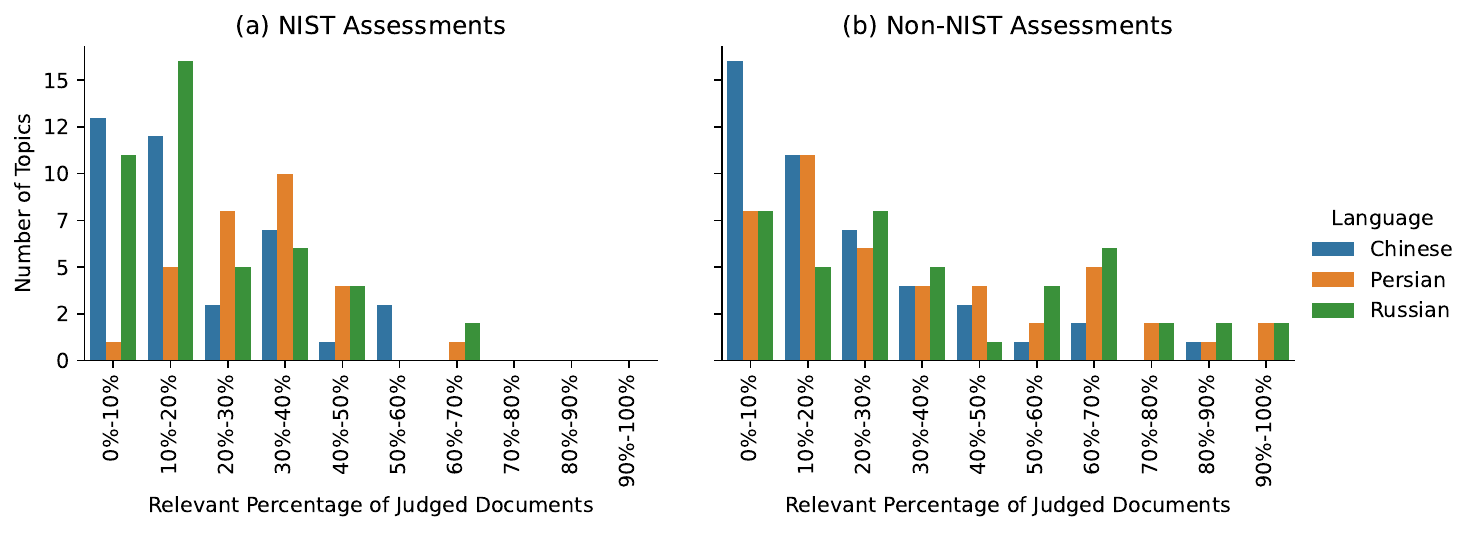}
\caption{Percentage of documents judged relevant reported by the two groups of assessors
during the preliminary topic development.}
\label{fig:topic_assess}
\end{figure*}

The viability of each topic in each language was determined using both NIST assessments and non-NIST assessments.
Figure~\ref{fig:topic_assess} shows the percentage of relevant documents found for annotated topics during initial development
by the two teams of assessors.
A topic was viable if at least one relevant document was found by a NIST or non-NIST assessor
and the topic did not appear to have too many relevant documents to support collection reusability.
If there was a disagreement about the prevalence ({\it{i.e}, percentage}) of relevant documents for a topic,
the NIST assessors value was used.
Therefore there were cases when the NIST assessor found at least one relevant document and the additional assessment did not find any.
In addition, there were topics where NIST assessors identified fewer than twenty relevant documents,
but the non-NIST assessors found that more than 65\% were relevant.
Disagreements could come from the fact that different documents were examined
and because relevance is an opinion.

\begin{figure*}[tb]
\centering
\includegraphics[width=\linewidth]{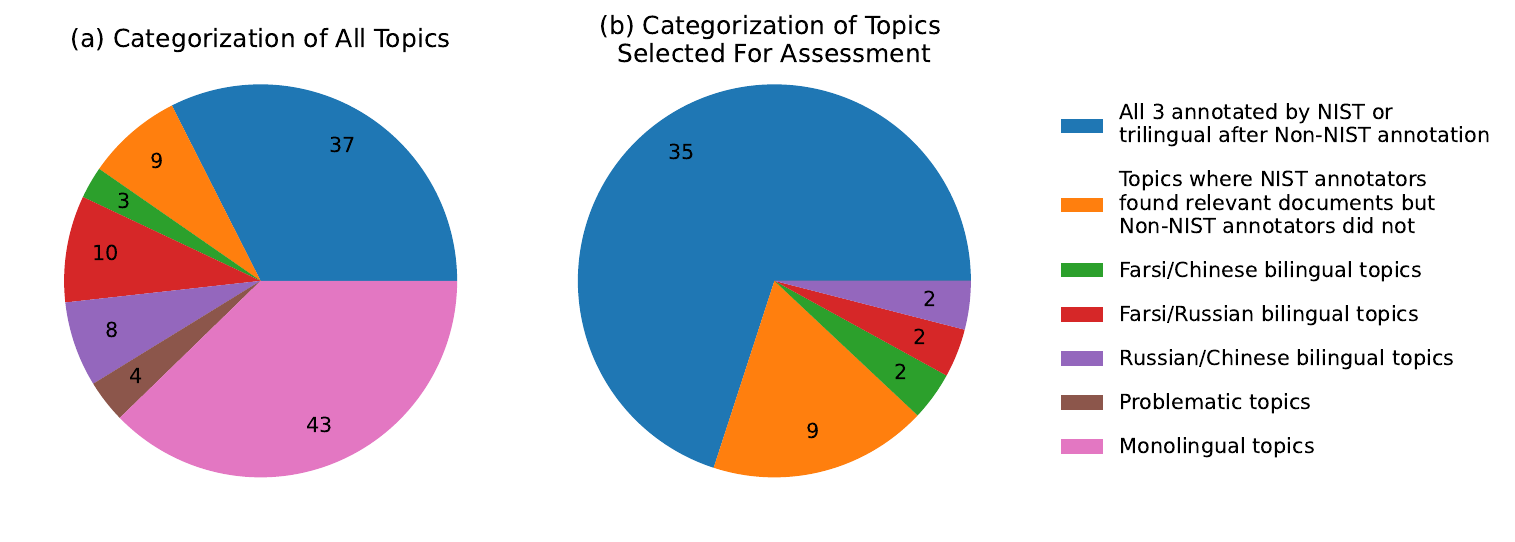}
\vspace{-3em}
\caption{Prioritization categories for all topics and for topics selected for pooling.}
\label{fig:topic_cat}
\end{figure*}

A priority list of topics was assembled to favor topics that appeared to be viable in all three languages
over topics that appeared to be viable in only two languages.
Each topic was assigned to a category.
Figure~\ref{fig:topic_cat} shows the distribution for all 114 topics
and the distribution of the fifty topics selected for further relevance assessment.
The intent was to evaluate all fifty topics in all three languages.

%% file: 5-judgments.tex
\section{Relevance Judgments}
\label{sec:rel-judge}

Once systems were run, relevance assessment began on the chosen subset of the 114 topics submitted for each run.
In the following we describe how the judgment pools were assembled and how relevance was determined.

\subsection{Creating Judgment Pools}

Pools were created from the top-ranked documents of submitting systems. 
The number of documents a run contributed to a pool
was based on whether the submitting team marked the run as a baseline run.
Baseline runs contributed their top twenty-five documents,
while non-baseline runs contributed their top fifty documents.
Thus for nDCG@20 all submissions have complete judgments.

\subsection{Relevance Judgment Process}

Assessors used a four-point scale to judge the relevance of each document.
Assessors were instructed to provide their assessment as if they were gathering information to write a report on the topic.
Relevance was assessed on the most valuable information found in the document;
the grades were:
\begin{description}
\itemsep0em 
\item[Very Valuable:] information that would appear in the lead paragraph of a report on the topic
\item[Somewhat Valuable:] information that would appear somewhere in a report on the topic
\item[Not that Valuable:] information that does not add new information beyond the topic description,
or information that would appear only in a report footnote
\item[Not Relevant:] a document without any information about the topic
\end{description}
The qrels use a three-point graded relevance scale:\footnote{An early release of the qrels had 3-2-1-0 graded relevance judgments corresponding to the 4-point scale used by assessors.}

\begin{description}
\item[3 points]  Very Valuable
\item[1 point] Somewhat Valuable
\item[0 points] Not that Valuable or Not Relevant
\end{description}

\input{_tab_label_confusion.tex}

\subsection{Analysis of Topics}

During the assessment period, forty-seven Chinese topics, forty-five Persian topics,
and forty-four Russian topics were judged.
Within each language some topics had fewer than three relevant documents,
while other topics had a concerningly large percentage of relevant documents in the pools.
Having topics with fewer than three relevant document can have undesirable effects on the ability to statistically distinguish systems.
There are three Chinese topics, four Persian topics, and three Russian topics with fewer than three relevant documents. 
Thus each language has at least forty topics; such is generally thought to be the minimum number of topics 
necessary to fairly compare systems.

Identifying topics with a large number of unidentified relevant documents is more important for future
use of the collection than for track participants,
since every research system had its top fifty documents judged.
Scores for systems are comparable and thus systems can be ranked using them. 
However, given the desire to create reusable collections,
determining topics that likely have many unjudged relevant documents is important.
One approach simply calculates the percentage of relevant documents in the pool
and sets a cutoff (such as 10\% prevalence)
as too high to be confident that the relevant set is sufficiently identified.
Using this cutoff would flag ten topics in Chinese, seven topics in Persian, and eleven topics in Russian.

\input{_fig_auto_depth}

Figure~\ref{fig:auto_depth} presents a closer investigation into the percentage of additional relevant documents found by non-baseline runs at various pool depths
(the \textit{gain curve}) by the automatic runs beyond those already found by depth 25 by some baseline run.
We exclude topics that do not discover more relevant documents as the depth increases (e.g., the baseline pool contains the same set of relevant documents as any automatic pool up to depth 50). 
We use a knee detection heuristic~\cite{cormack2016engineering} on the gain curves to identify topics that are less likely to find a sizable number of unjudged documents with a deeper pool.
We calculate the maximum log slope ratio over any automatic pool depth $d$ as the indicator of curvature.
Specifically, let $slope_{i, j}$ be the slope of the curve from $i$ to $j$, i.e., $(p_j-p_i)/(j-i)$ where $p_k$ is the percentage of the relevant documents found with pool depth $k$.
The maximum log slope ratio is defined as:
\begin{equation*}
    \max_{d\in [1, 50]} \log \left( \frac{ slope_{1, d} }{ slope_{d, 50} }  \right).
\end{equation*}
To capture the general trend of the curve instead of sharp fluctuations,
we smooth the curve by averaging a moving window of size three. 
If the value is 0, the curve is a straight line without a plateau;
a log ratio that is close to infinity indicates a flat line toward the end. 

As demonstrated in Figure~\ref{fig:auto_depth}, topics with higher prevalence tend to be continuously finding more relevant documents as the pools become larger, i.e., less likely to be completely judged, which aligns with the prior heuristics on assessing the completeness by prevalence. 
Topics with moderate prevalence tend to be more complete but not guaranteed. However, the range of appropriate prevalence is subject to the language, and potentially the collection and participating systems. 
Nevertheless, these results these results suggest some considerations that we will bear in mind during topic curation next year. 

\subsection{Assessor Agreement}\label{sec:iaa}

\begin{table}[b]
\centering
\caption{Cohen's $\kappa$ assessor agreement on a sample of relevance assessments, by language. $\kappa$ values are annotated with interpretations~\cite{viera2005understanding} of (F)air, (M)oderate, and (S)ubstantial agreement (others are slight).}
\label{tab:kappa}
\begin{tabular}{lrrrr}
\toprule
Labels & Chinese & Persian & Russian & Overall \\
\midrule
4 labels & (M) 0.515 &  0.081 & (F) 0.300 & (F) 0.346 \\
3 labels & (F) 0.376 & 0.131 & (M) 0.460 & (F) 0.392 \\
Binary & (M) 0.557 &  0.151 & (M) 0.541 & (M) 0.524 \\
Fuzzy & (S) 0.777 & (F) 0.326 & (M) 0.591 & (S) 0.674 \\
\bottomrule
\end{tabular}
\end{table}

As is the convention with TREC tracks, the official relevance assessments for a given topic represent a single assessor's interpretation of the topic and assessment guidelines.
Nevertheless, we sought to explore whether different assessors would have determined different documents to be relevant for the topic
and whether such user differences would affect the overall system rankings.
We explore the former question in this section and the latter question in Section~\ref{sec:assessor_corr}.

An additional NIST assessor labeled the relevance of the pooled documents for 28 topic-language pairs
(12 topics in Chinese, 8 topics in Persian, and 8 topics in Russian),
using the same assessment criteria that the topic's official assessor used. 

Table~\ref{tab:label_confusion} presents confusion matrices for each language.
We observe that a high proportion of the differences in labels are between the \textit{Not that Valuable} (NV) and the \textit{Not Relevant} (NR) labels,
87\% for Persian, 66\% for Russian, and 41\% for Chinese.
These results motivate our decision to merge these labels during evaluation.

Given the unbalanced nature of the relevance labels,
we compute the Cohen's $\kappa$ coefficient~\cite{cohen1960coefficient} to assess the overall agreement of the labels.
We explore agreement in four settings:
the original 4 relevance labels;
the 3 labels used for evaluation (merging \textit{Not that Valuable} and \textit{Not Relevant});
binary labels used for evaluation measures like MAP (further merging \textit{Very Valuable} and \textit{Somewhat Valuable});
and a ``Fuzzy'' setting, in which adjacent labels are considered matches.
Table~\ref{tab:kappa} presents the $\kappa$ values for each of these settings
alongside the established interpretations of their quality from \citet{viera2005understanding}.
We find that agreement improves for Persian and Chinese when the bottom two relevance labels are merged,
and further improves for all languages in the binary setting.
When we consider adjacent relevance scores as matches (the Fuzzy setting),
we observe substantial agreement in Chinese, moderate agreement in Russian, and fair agreement in Persian.
These results suggest that the Persian relevance labels may be biased towards the topic's specific assessor,
while the Chinese and Russian labels potentially generalize better across users.
Further, while there can be disagreement among assessors about the exact degree of relevance,
such cases are generally among adjacent labels.

\section{Additional Resources}

The track provided three types of additional resources: translated documents, translated queries, and translations of MS MARCO into the collection languages 
(NeuMARCO\footnote{\url{https://huggingface.co/datasets/neuclir/neumarco}}).  
Links to some other types of pre-existing resources that might be useful to participants were also provided.

One additional resource the track provided was machine translations of the documents into English
and the queries into Chinese, Persian, and Russian.
These resources facilitated meaningful comparison across systems that used machine translation to cross the language barrier.
Documents were translated using a vanilla Transformer model that was trained in-house with the \textit{Sockeye} version 2 toolkit \cite{hical} using bitext obtained from publicly available corpora.\footnote{\url{https://opus.nlpl.eu}}
The number of sentences used in training is given in Table \ref{tbl:bitext}, along with BLEU scores on the FLORES-101 benchmark \cite{goyal-etal-2022-flores} for each language.

\begin{table}
\centering
\caption{MT Training Data}
\label{tbl:bitext}
\begin{tabular}{lrr}
\toprule
Language & $\#$ Sentences & BLEU \\
\midrule
Chinese & 84,764,463 & 31.5 \\
Persian & 11,426,143 & 35.1 \\
Russian & 119,856,685 & 34.9 \\
\bottomrule
\end{tabular}
\vspace{-0.5em}
\end{table}

Query translations were obtained from \textit{Google Translate} since its performance on titles was superior.
While no team processed their own translations of the documents,
one team produced their own translations of the queries. 

The track website also collected a number of multilingual and bilingual resources in the languages of the track
including translations of MSMARCO passages into the document languages~\cite{nonifacio-mMarco}; 
HC4 – a CLIR collection built over three years of Common Crawl data in the same three languages~\cite{hc4};
and two multilingual CLIR datasets based off of Wikipedia known as CLIRMatrix~\cite{clirmatrix} and WikiCLIR~\cite{wikiclir}.

%% file: _tab_label_confusion.tex
\begin{table*}[t]
\centering
\caption{Inter-assessor confusion matrices, as raw label counts. Key: \textit{Very Valuable} (VV),  \textit{Somewhat Valuable} (SV), \textit{Not that Valuable} (NV), \textit{Not Relevant} (NR).}
\label{tab:label_confusion}

\begin{tabular}{llrrrr}
\multicolumn{6}{c}{\bf Chinese} \\
\toprule
&&\multicolumn{4}{c}{Second Label} \\
& & NR & NV & SV & VV \\
\cmidrule(lr){3-6}
\multirow{4}{*}{\rotatebox[origin=c]{90}{Official}}
& \multicolumn{1}{l|}{NR} & 7694 & 260 & 22 & 0 \\
& \multicolumn{1}{l|}{NV} & 165 & 291 & 71 & 3 \\
& \multicolumn{1}{l|}{SV} & 49 & 155 & 94 & 1 \\
& \multicolumn{1}{l|}{VV} & 34 & 99 & 185 & 20 \\
\bottomrule
\end{tabular}
\hspace{1.5em}
\begin{tabular}{llrrrr}
\multicolumn{6}{c}{\bf Persian} \\
\toprule
&&\multicolumn{4}{c}{Second Label} \\
& & NR & NV & SV & VV \\
\cmidrule(lr){3-6}
\multirow{4}{*}{\rotatebox[origin=c]{90}{Official}}
& \multicolumn{1}{l|}{NR} & 4996 & 18 & 7 & 3 \\
& \multicolumn{1}{l|}{NV} & 804 & 25 & 19 & 7 \\
& \multicolumn{1}{l|}{SV} & 54 & 8 & 4 & 3 \\
& \multicolumn{1}{l|}{VV} & 12 & 5 & 0 & 4 \\
\bottomrule
\end{tabular}
\hspace{1.5em}
\begin{tabular}{llrrrr}
\multicolumn{6}{c}{\bf Russian} \\
\toprule
&&\multicolumn{4}{c}{Second Label} \\
& & NR & NV & SV & VV \\
\cmidrule(lr){3-6}
\multirow{4}{*}{\rotatebox[origin=c]{90}{Official}}
& \multicolumn{1}{l|}{NR} & 4854 & 59 & 17 & 17 \\
& \multicolumn{1}{l|}{NV} & 620 & 53 & 27 & 71 \\
& \multicolumn{1}{l|}{SV} & 96 & 10 & 10 & 44 \\
& \multicolumn{1}{l|}{VV} & 44 & 13 & 13 & 126 \\
\bottomrule
\end{tabular}

\end{table*}

%% file: _fig_auto_depth.tex
\begin{figure*}
    \centering
    \includegraphics[width=0.9\linewidth]{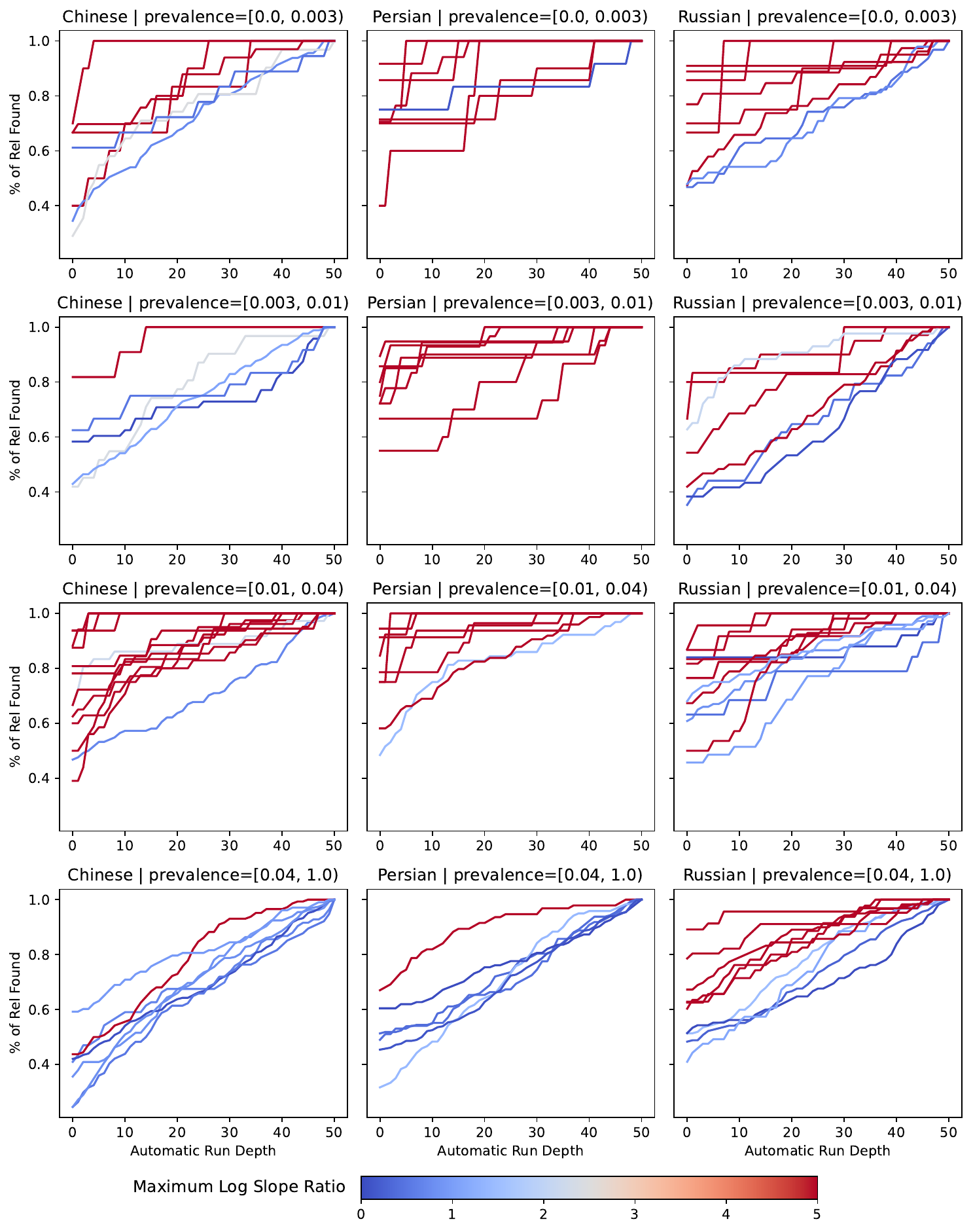}
    \caption{Percentage of relevant documents found with different pooling depths on the runs. The depth for the baseline runs is set to 25. Topics are grouped by prevalence, which is the percentage of the relevant document found among all judged documents. }
    \label{fig:auto_depth}
\end{figure*}

%% file: 6-participation.tex
\section{Participants}

Including baselines, we scored 52 Chinese runs, 60 Persian runs, and 60 Russian runs.  
Table~\ref{tab:task-cnts} outlines the number of runs submitted to each of the tasks: monolingual IR, ad hoc CLIR, and reranking CLIR. 
A total of 12 teams submitted runs for at least one language. This track had worldwide
participation, with three teams from Asia,
one from Europe, one from South America, and the remainder from North America. More information about participant systems is available in the teams' notebook papers.

\begin{table}
\centering
\caption{Number of runs submitted to each task}
\label{tab:task-cnts}
\begin{tabular}{lrrrr}
\toprule
Language & Monolingual & Ad Hoc & Reranking & Total \\
\midrule
Chinese & 17 & 30 & 5 & 52 \\
Persian & 20 & 34 & 6 & 60 \\
Russian & 20 & 35 & 5 & 60 \\
\bottomrule
\end{tabular}
\end{table}

%% file: 7-results.tex
\begin{table*}
\caption{Average effectiveness by the type of the CLIR runs. }\label{tab:results_agg_by_model}
\centering
\begin{tabular}{l|ccc|ccc|ccc}
\toprule
               & \multicolumn{3}{c|}{Chinese} & \multicolumn{3}{c|}{Persian} & \multicolumn{3}{c}{Russian} \\
               &    nDCG &    MAP &   R@1k &    nDCG &    MAP &   R@1k &    nDCG &    MAP &   R@1k \\
\midrule
Rerank         &   0.299 &  0.218 &  0.781 &   0.391 &  0.267 &  0.817 &   0.376 &  0.263 &  0.774 \\
\midrule
Hybrid         &   0.419 &  0.282 &  0.695 &   0.439 &  0.313 &  0.788 &   0.516 &  0.396 &  0.800 \\
Dense          &   0.199 &  0.131 &  0.463 &   0.198 &  0.123 &  0.497 &   0.224 &  0.143 &  0.496 \\
Learned-sparse &     --  &    --  &    --  &   0.449 &  0.300 &  0.834 &   0.437 &  0.321 &  0.791 \\
Sparse         &   0.283 &  0.207 &  0.657 &   0.290 &  0.195 &  0.712 &   0.294 &  0.212 &  0.679 \\
\bottomrule
\end{tabular}
\vspace{2em}
\end{table*}

\section{Results and analysis}

In this section, we summarize the results and provide some analysis on topic difficulty, reusability, and the effect on system preference order of using different annotators.

\subsection{Overall Results}

The full results are presented in Tables~\ref{tab:zho-full-results}, \ref{tab:fas-full-results}, and \ref{tab:rus-full-results}. The top-ranked systems all use a combination of title and description queries. 
Table~\ref{tab:results_agg_by_model} summarizes the effectiveness of systems categorized by the type of the model.
Since huaweimtl team indicates that runs \texttt{huaweimtl-\{zh,fa,ru\}-m-hybrid1} runs were ensembling systems that includes a monolingual system (i.e., using human translated queries), these three runs are marked as monolingual runs by the organizers.

On average, hybrid approaches that combine dense and sparse retrieval in the system tend to provide the best nDCG@20. 
Both hybrid and learned-sparse (such as SPLADE~\cite{splade}) models provide a recall at 1000 close to 80\%. 
Note that the reranking runs tend to have a higher recall at 1000, which is based on a BM25 retrieval result with document translation, that should not be attributed to the reranking models. 

The variation among dense retrieval models is large, as we can observe in Figure~\ref{fig:ndcg-bar}. 
Several dense retrieval models are among the top-ranked systems while others are scattered throughout their peers. 
Sparse retrieval systems provide a moderate performance, which is mostly contributed by the baseline runs, which are contributed by the organizers and several participants~\cite{lin2023neuclir-baselines}. 

\input{_fig_ndcg_bar.tex}

The left column of Figure~\ref{fig:ndcg-bar} presents the monolingual runs. 
Despite not being the focus of the NeuCLIR track, they enrich the pool and provide a target for the CLIR models to compare with. 
The highest performing CLIR system, in terms of nDCG@20, for Chinese and Persian outperformed the monolingual model for the corresponding language by about 0.015; the best Russian CLIR system achieved about the same nDCG@20 as the best Russian monolingual system. 
We defer the discussion on the pooling enrichment benefit of the monolingual runs to Section~\ref{sec:reusability}. 

\subsection{Topic Difficulty}

\input{_fig_tsne.tex}

One of the objectives of topic development is to create a set of topics where the retrieval results are able to distinguish systems.
Topics that are too easy or not having any relevant documents are not ideal. 
Figures~\ref{fig:topic-box-zho}, \ref{fig:topic-box-fas}, and \ref{fig:topic-box-rus} are nDCG@20 boxplots of all the judged topics for each language. 
Topic 118 for Persian is an example of an easy topic where 75\% of the runs achieve nDCG@20 over 0.80; 
in contrast, all runs score zero for topics 4 and 24 against the Chinese documents, indicating that these two topics are not likely to have any relevant document in the collection.
In a practical sense, when no run has retrieved any relevant document, no relevant documents are judged during pooling, thus the topic is not usable for evaluating future systems. 
Topics, such as Topic 24 in Russian, with a wide range of scores, are ideal for distinguishing systems. 

However, topics such as 52 in Chinese and 4 in Persian can give future systems that can understand the complex semantics of the queries an advantage, and, therefore, reflect the systems' improvement. 
Although most systems have low nDCG scores for these topics, there are relevant documents judged thanks to pooling. 
Future systems that are able to retrieve these rare but relevant documents will be able to obtain a higher overall score. 

Systems retrieved more relevant documents for topics that are not related to any country or region, such as Topic 32 (\textit{Peanut allergy treatment}) and 16 (\textit{Dramatic decrease in number of bees}). 
Topics that are more specific to the country where the language is widely spoken tend to result in retrieval of larger numbers of relevant documents. 
For example, Topic 4 (\textit{Corruption during construction of Vostochny Cosmodrome}) is among the easiest topics for Russian; however, there are no relevant documents in Chinese, and it is extremely challenging for Persian. 
Such topics with disjoint interests in different languages are not particularly problematic for evaluating CLIR but this could be an important issue in future MLIR evaluations in which the document collection contains multiple languages. 

\subsection{Retrieval Diversity}
\input{_fig_lou_scatter.tex}

Forming a diverse set of judgments could lead to a more reusable collection for evaluating future systems, and such judgments require a diverse set of retrieval results. 
For each run, we form a vector with the nDCG@20 values of each topic. Thus, the size of such vectors is the number of the judged topics in the language.
Figure~\ref{fig:tsne} plots tSNE graphs that project the nDCG@20 vectors to two dimensions. The shade of the markers indicates the overall nDCG@20 of the run. 

Among different run types (Figure~\ref{fig:tsne}(a)), there is not a clear cluster that gathers monolingual systems, which indicates that the monolingual subtask might not provide much value for diversifying the pool. 
End-to-end CLIR systems (i.e., no translation involved during inference\footnote{We inferred this from the submission metadata and considered systems marked using English queries and native documents as end-to-end CLIR runs.}) 
demonstrate two clear clusters in each language, while having a clear separation between runs that involve query translation. 

There are more separations among the model types. 
Hybrid runs cluster together in all languages, with a clear distance from the sparse runs. 
Several dense runs are among runs with high overall scores~(darker colors), while others are among the lowest-scoring runs. 
This indicates that the trend appears in not only the overall scores (Figure~\ref{fig:ndcg-bar}) but also behavior on individual topics. 

Figures~\ref{fig:zho-overlap}, \ref{fig:fas-overlap}, and \ref{fig:rus-overlap} plot the retrieval similarity among all submissions, where lighter colors indicate higher similarity. 

For the top 100 retrieved documents (Figures \ref{fig:zho-overlap}(a), \ref{fig:fas-overlap}(a), and \ref{fig:rus-overlap}(a)), runs submitted by the same team tend to be more similar than others, which might indicate that teams are often submitting ablated runs instead of building different stacks of systems. 
Top-ranked systems also tend to be more similar to each other as they all put relevant documents at the top, with a more clear trend in Persian and Russian than in Chinese. 
Sparse runs also retrieve highly similar sets of documents in the top 100, especially in Persian and Russian, indicating that the ranking model might all be leveraging similar features. 

For the retrieved relevant documents, the trend is similar, with top-ranked runs demonstrating higher similarity. 
However, the light triangle in the middle is less clear, indicating that despite providing lower recall, these runs still contribute some unique relevant documents to make the pool more diverse, which leads to a more reusable collection in the future. Note that the numerator and the denominator of the similarity are the sizes of the intersection and the union of the two runs; this similarity measure would give low similarity if the recall of the two runs are too different. 

\subsection{Reusability}\label{sec:reusability}

To evaluate the reusability of the collection, we conduct the Leave-Out-Unique experiment~\cite{buckley2007bias, zobel1998reliable} to test the robustness of the collection. 
In the experiment, we leave out one run from the construction of the pool and evaluate the run with the modified qrels. This process simulates the possibility of each run being a future run that does not participate in pooling. 
Since the primary purpose of the evaluation is to rank systems, the differences in the actual values of the evaluation metric after modifying the qrels are negligible if the ordering of the runs remains the same. Therefore, we calculate Kendall's $\tau$ on the rank of the systems for quantitatively evaluating the correlation. 
Figure~\ref{fig:lou-scatter} demonstrate the experiment results. Each dot indicates a run where the x-axis indicates the evaluation metric on the full qrels and the y-axis is the modified version. 
The results illustrate that most of the runs are still on the diagonal, indicating that the collection is indeed robust and can fairly evaluate future systems that do not contribute to the pools with Kendall's $\tau$ close to 0.99. 

Since the variation between the runs submitted by a team is small, we additionally conduct a Leave-Out-Team-Unique experiment where we remove all submissions from the same team when evaluating a run. Such experiments are more aggressive than the Leave-Out-Unique experiments but provide a more adequate assessment of the reusability. 
Figure~\ref{fig:lotu-scatter} presents the results. The correlation between the full pool and the modified pool is lower with 0.98 for nDCG@20 and 0.95 for recall at 1000. However, we argue that the correlation is still high enough for fairly evaluating future systems. 

\subsection{Assessor Effect on System Preference Order}\label{sec:assessor_corr}

\begin{table}[tb]
\centering
\caption{Correlation between systems when measured using the official assessments and the second assessor's labels. All correlations are significant at $p<0.001$.}
\label{tab:rank_corr}
\begin{tabular}{llrr}
\toprule
Language & Measure & Spearman's $\rho$ & Kendall's $\tau$ \\
\midrule
\multirow{5}{*}{Chinese}
& nDCG@20 & 0.951 & 0.829 \\
& RBP & 0.960 & 0.843 \\
& MAP & 0.975 & 0.883 \\
& R@100 & 0.971 & 0.881 \\
& R@1000 & 0.960 & 0.846 \\
\midrule
\multirow{5}{*}{Persian}
& nDCG@20 & 0.829 & 0.650 \\
& RBP & 0.818 & 0.628 \\
& MAP & 0.831 & 0.652 \\
& R@100 & 0.687 & 0.509 \\
& R@1000 & 0.756 & 0.575 \\
\midrule
\multirow{5}{*}{Russian}
& nDCG@20 & 0.928 & 0.777 \\
& RBP & 0.878 & 0.708 \\
& MAP & 0.894 & 0.728 \\
& R@100 & 0.732 & 0.561 \\
& R@1000 & 0.708 & 0.544 \\
\bottomrule
\end{tabular}
\end{table}

Given the variable agreement levels among assessors observed in Section~\ref{sec:iaa},
we explore whether the ranking of submitted systems differs when using the alternative assessments.
To this end, we compare the nDCG@20, RBP, MAP, R@100, and R@1000 of the systems using official judgments and the second judgments
over the 28 re-assessed topic-language pairs.
We measure the rank correlation of the systems using Spearman's $\rho$ and Kendall's $\tau$ statistics,
and present the results in Table~\ref{tab:rank_corr}.
We observe a very strong correlation for nDCG@20, RBP, and MAP ($\rho>0.83$ and $\tau>0.62$)
and a strong correlation for R@100 and R@1000 ($\rho>0.68$ and $\tau>0.50$).
Noting that using only 28 topics in this analysis induces a greater degree of random variation than would be the case for the full topic set, these results suggest that although assessors sometimes disagree on relevance labels,
the system preference order may not change much if a different assessor provided the labels.

%% file: _fig_ndcg_bar.tex
\begin{figure*}
    \centering
    \includegraphics[width=0.95 \linewidth]{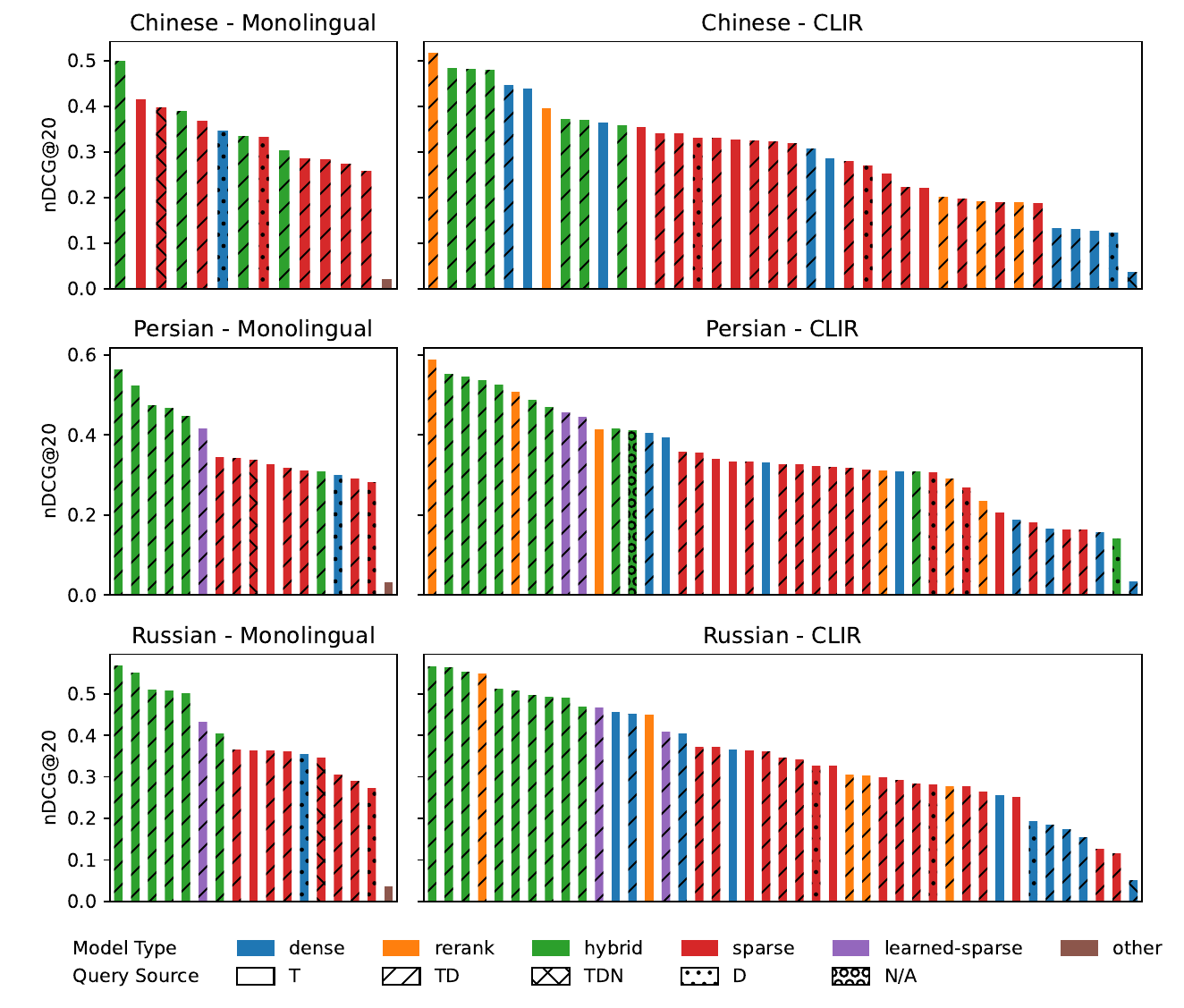}
\caption{Bar charts of submitted runs. Monolingual and CLIR runs are separated into subplots for clarity.}\label{fig:ndcg-bar}
\end{figure*}

%% file: _fig_tsne.tex
\begin{figure*}
    \centering
    \includegraphics[width=\linewidth]{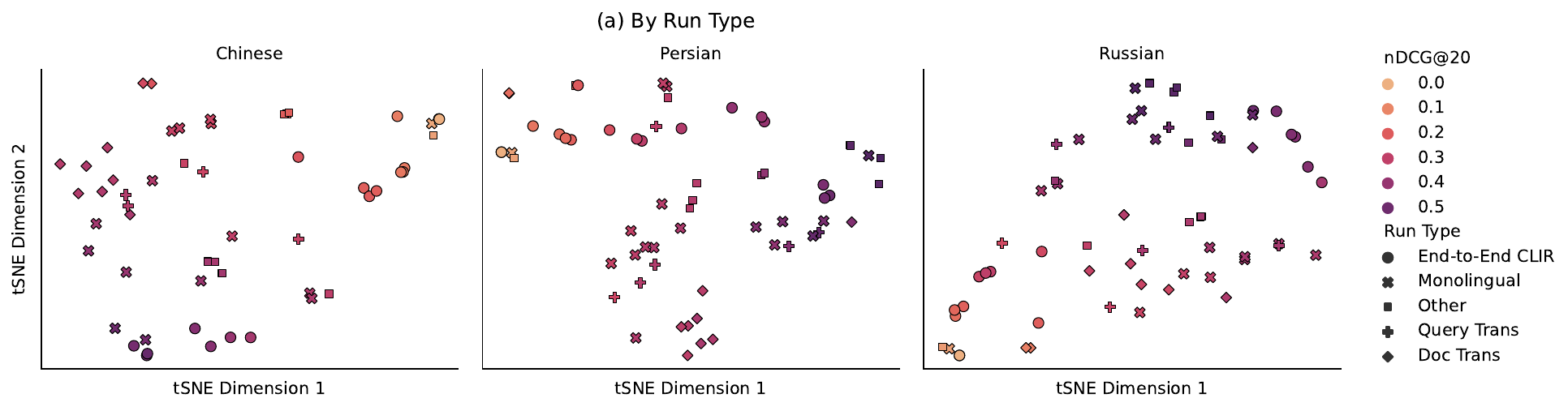}
    \includegraphics[width=\linewidth]{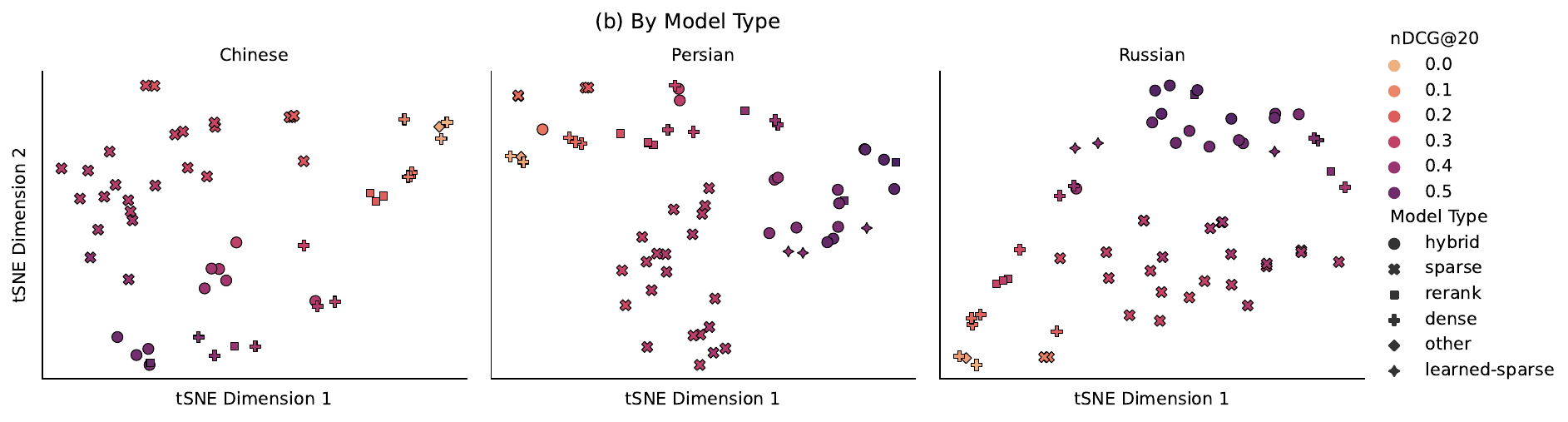}
    \caption{tSNE graphs of nDCG@20 for each submitted run. The shade of the marker indicates the overall nDCG@20 of the run. }
    \label{fig:tsne}
\end{figure*}

%% file: _fig_lou_scatter.tex
\begin{figure*}
    \centering
    \includegraphics[width=0.85\linewidth]{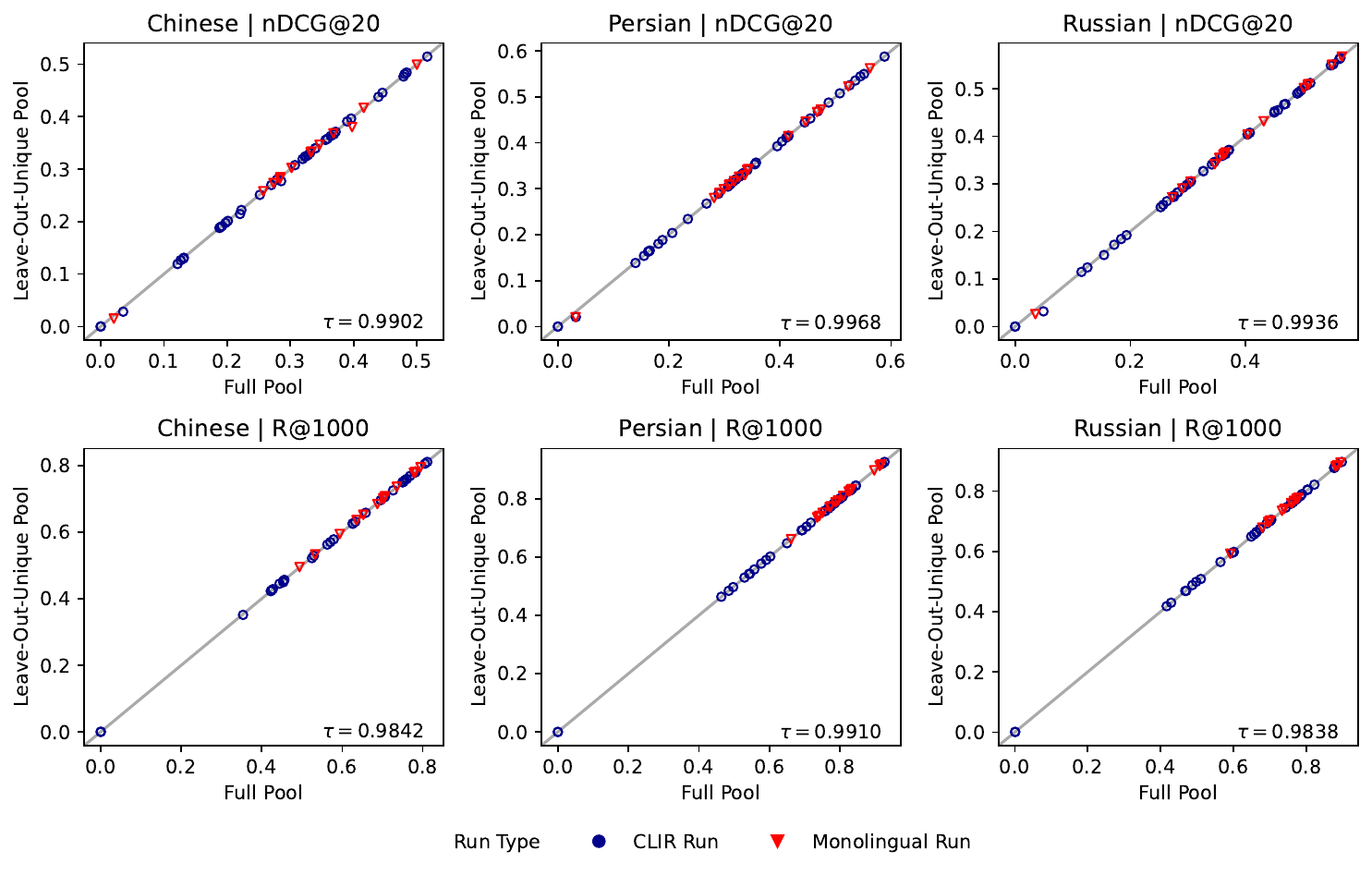}
    \caption{Leave-Out-Unique pool experiments. Rank correlations measured by Kendall's $\tau$ are marked at the corner of each graph. }
    \label{fig:lou-scatter}
\end{figure*}

\begin{figure*}
    \centering
    \includegraphics[width=0.85\linewidth]{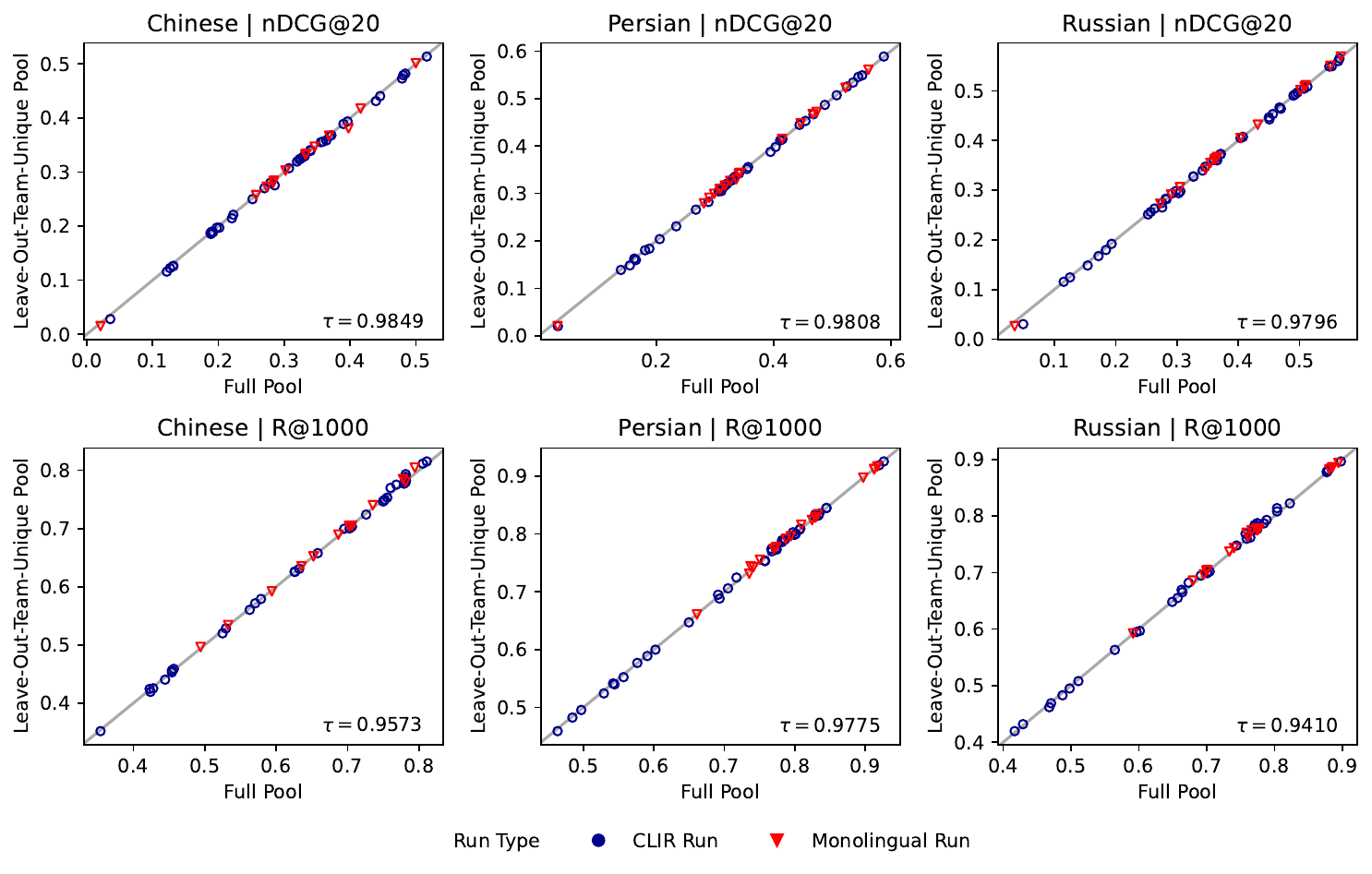}
    \caption{Leave-Out-Team-Unique pool experiments. Rank correlations measured by Kendall's $\tau$ are marked at the corner of each graph.}
    \label{fig:lotu-scatter}
\end{figure*}

%% file: 8-future.tex
\section{Future Track Directions}

The TREC NeuCLIR track will continue in 2023.
The CLIR task will continue, with new topics for the same collections in the same three languages.
In this way, we hope to support both improved designs and improved training,
since training data from this first year of the track will be available
with characteristics closely approximating those of the 2023 CLIR task.  

We are considering three additional tasks:
\begin{itemize}
\item Multilingual Information Retrieval (MLIR).
In this task, systems would be asked to create a single ranked list over all three test collection languages (Chinese, Persian and Russian),
and would be evaluated using the relevance judgments for all three languages.
A similar MLIR task was evaluated in the Cross-Language Evaluation Forum (CLEF) and found to be challenging~\cite{ferro2019information}.
\item Non-English Queries (NEQ).
In this task, systems would perform CLIR or MLIR using queries that are not in English.
The present NeuCLIR test collection can support NEQ tasks with Chinese, Persian or Russian queries,
and generation of topics in a small number of additional languages may also be possible. 
\item CLIR for Technical Documents (CLIR-TD).
CLIR could be of value in many domain-specific applications, including for example law, medicine, or engineering.  
As a first step toward evaluating domain-specific applications of CLIR we are considering the addition of a pilot task in which the goal is to perform CLIR for technical documents in a single non-English language (either Chinese, Persian or Russian) using English queries.
A promising direction is to use repositories of non English academic text, such as CNKI\footnote{https://www.cnki.net} or Wanfang\footnote{\url{http://www.wanfangdata.com}} for Chinese, or the Russian Science Citation Index\footnote{\url{https://elibrary.ru}} for Russian, or to select dual-language biomedical content that is indexed in PubMed. 
\end{itemize}

Starting three new tasks in the same year would risk dividing participating research teams in ways that would work against our goal of being able to compare results and foster progress on specific tasks, however.  For this reason, we want to use our planning session at TREC 2022 to discuss at least these four questions with participants:
\begin{itemize}
  \item What changes to the CLIR task would further enhance the ability of participating teams to achieve their research goals?
  \item Which of the additional tasks we have listed are of greatest interest to participants (and thus most likely to attract a critical mass of participation)?
  \item Are there other new tasks that we should be considering that might be an even higher priority for NeuCLIR in 2023?
  \item Are there additional outreach opportunities that might attract a substantial number of new participants, and if so which new tasks might be most likely to help attract those new participants?
\end{itemize}

We also plan to re-evaluate tools and methodology for reporting carbon footprint of participating systems. 
We note that, in 2022, only three teams reported their energy usage;
many cited the lack of tools compatible with their system,
while others said that keeping track of all stages of a track submission is too onerous and error prone
(e.g., one might forget to log an experiment). 
Further, some ambiguity existed in how external resources should be accounted for;
for example, should energy required to run systems that are shared with other projects in a lab be counted? 
Possible improvements in energy tracking include:
\begin{itemize}
    \item Ask participants to measure impact only while preparing their submission: this would exclude any energy used during the training of neural models or indexing of document collections. 
    Organizers could still use the information collected during a run submission to estimate the total energy usage.
    \item Explicitly formalize alternative metrics to energy reporting: this could include compute-hours, hardware, or other information that could be used to estimate energy impact.
    \item Defer energy reporting from run submission time to notebook submission time: this would give teams more time to analyze their impact without having to sacrifice time in the weeks immediately preceding runs submission.
\end{itemize}

%% file: 9-conclusion.tex
\section{Conclusion}

CLIR at TREC is back!
In this first year of NeuCLIR we have worked together to forge a research community,
create new evaluation resources,
and generate what is for the moment the world's largest collection of neural CLIR results
that have been run under comparable conditions.
Twelve participating teams (and several baseline systems) contributed a total of 172 runs,
achieving strong first-year results that substantially outperformed non-neural baselines.
Moreover, we might reasonably expect the 2022 NeuCLIR relevance judgments
that are now available to future participants to offer the potential for continued improvement.
The NeuCLIR track will continue at TREC 2023,
perhaps with one or more additional tasks, so we have much to look forward to.

%% file: _tab-full-results-zho.tex
\begin{table*}[]
\caption{Chinese run results. Monolingual runs, which use human translations of the queries, are marked as green.
Run used as the first stage retrieval for the reranking task is marked bold. 
* indicates manual runs.
}\label{tab:zho-full-results}
\centering
\resizebox{\linewidth}{!}{
\begin{tabular}{ll|ccccccc|cccccc}
\toprule
       Team & Run ID &  Man. & ReR. &  QF &  QL & DL & Model &    PRF &  nDCG & RBP &    MAP & R@100 & R@1k \\
\midrule
  NM.unicamp\cite{participants-NM.unicamp} &                   p2.zh.rerank & \xmark &    \cmark &          TD &    English &   Native &      Dense & \xmark &   0.516 &      0.553 & 0.404 & 0.591 &  0.781 \\
  NM.unicamp\cite{participants-NM.unicamp} &                \ml{p3.zh.mono} & \xmark &    \xmark &          TD &      Human &   Native &     Hybrid & \cmark &   0.500 &      0.532 & 0.384 & 0.584 &  0.702 \\
   CFDA\_CLIP\cite{participants-CFDA-CLIP} &            CFDA\_CLIP\_zho\_dq & \xmark &    \xmark &          TD &    English &   Native &     Hybrid & \xmark &   0.484 &      0.528 & 0.360 & 0.549 &  0.750 \\
   CFDA\_CLIP\cite{participants-CFDA-CLIP} &             CFDA\_CLIP\_zho\_L & \xmark &    \xmark &          TD &    English &   Native &     Hybrid & \xmark &   0.481 &      0.522 & 0.345 & 0.577 &  0.750 \\
   CFDA\_CLIP\cite{participants-CFDA-CLIP} &           CFDA\_CLIP\_zho\_clf & \xmark &    \xmark &          TD &    English &   Native &     Hybrid & \xmark &   0.479 &      0.527 & 0.349 & 0.566 &  0.750 \\
  hltcoe-jhu\cite{participants-hltcoe-jhu} &         coe22-tdq-zho\_colxtt* & \cmark &    \xmark &          TD &    English &   Native &      Dense & \xmark &   0.446 &      0.470 & 0.350 & 0.570 &  0.811 \\
  hltcoe-jhu\cite{participants-hltcoe-jhu} &          coe22-tq-zho\_colxtt* & \cmark &    \xmark &           T &    English &   Native &      Dense & \xmark &   0.439 &      0.465 & 0.332 & 0.542 &  0.781 \\
                                         - &      \ml{coe22-bm25-t-ht-zho*} & \cmark &    \xmark &           T &      Human &   Native &     Sparse & \cmark &   0.416 &      0.431 & 0.320 & 0.566 &  0.778 \\
                                         - &            \ml{coe22-man-zho*} & \cmark &    \xmark &         TDN &      Human &   Native &     Sparse & \xmark &   0.398 &      0.453 & 0.273 & 0.441 &  0.687 \\
            KASYS\cite{participants-KASYS} &     KASYS\_onemodel-rerank-zho & \xmark &    \cmark &           T &    English &   Native &      Dense & \xmark &   0.396 &      0.440 & 0.286 & 0.476 &  0.781 \\
    huaweimtl\cite{participants-huaweimtl} &    \ml{huaweimtl-zh-m-hybrid1} & \xmark &    \xmark &          TD &      Human &   Trans. &     Hybrid & \xmark &   0.390 &      0.397 & 0.255 & 0.498 &  0.756 \\
    huaweimtl\cite{participants-huaweimtl} &         huaweimtl-zh-c-hybrid3 & \xmark &    \xmark &          TD &      Other &   Trans. &     Hybrid & \xmark &   0.372 &      0.382 & 0.233 & 0.464 &  0.706 \\
  NM.unicamp\cite{participants-NM.unicamp} &                      p1.zh.hoc & \xmark &    \xmark &          TD &      Other &   Native &     Hybrid & \cmark &   0.369 &      0.420 & 0.218 & 0.370 &  0.454 \\
                                         - &     \ml{coe22-bm25-td-ht-zho*} & \cmark &    \xmark &          TD &      Human &   Native &     Sparse & \cmark &   0.368 &      0.405 & 0.296 & 0.517 &  0.794 \\
            KASYS\cite{participants-KASYS} &          KASYS\_one\_model-zho & \xmark &    \xmark &           T &    English &   Native &      Dense & \xmark &   0.364 &      0.395 & 0.222 & 0.367 &  0.563 \\
    huaweimtl\cite{participants-huaweimtl} &         huaweimtl-zh-c-hybrid2 & \xmark &    \xmark &          TD &      Other &   Trans. &     Hybrid & \xmark &   0.359 &      0.363 & 0.213 & 0.444 &  0.703 \\
                                         - &           coe22-bm25-t-dt-zho* & \cmark &    \xmark &           T &    English &   Trans. &     Sparse & \cmark &   0.356 &      0.376 & 0.281 & 0.503 &  0.805 \\
    huaweimtl\cite{participants-huaweimtl} &       \ml{zh\_xdpr.ms.oht.d.R} & \xmark &    \xmark &           D &      Human &   Native &      Dense & \cmark &   0.346 &      0.393 & 0.236 & 0.364 &  0.594 \\
                                         - & \textbf{coe22-bm25-td-dt-zho*} & \cmark &    \xmark &          TD &    English &   Trans. &     Sparse & \cmark &   0.340 &      0.360 & 0.264 & 0.502 &  0.781 \\
jhu.mcnamee\cite{participants-jhu.mcnamee} &                jhumc.zh5.td.rf & \xmark &    \xmark &          TD &    English &   Trans. &     Sparse & \cmark &   0.339 &      0.364 & 0.275 & 0.484 &  0.779 \\
                                    h2oloo &         \ml{zh\_dense-rrf.prf} & \xmark &    \xmark &          TD &      Human &   Native &     Hybrid & \cmark &   0.333 &      0.368 & 0.235 & 0.398 &  0.636 \\
                                         - &      \ml{coe22-bm25-d-ht-zho*} & \cmark &    \xmark &           D &      Human &   Native &     Sparse & \cmark &   0.332 &      0.356 & 0.266 & 0.476 &  0.735 \\
                                         - &           coe22-bm25-d-dt-zho* & \cmark &    \xmark &           D &    English &   Trans. &     Sparse & \cmark &   0.331 &      0.328 & 0.254 & 0.482 &  0.760 \\
                                         - &          coe22-bm25-td-mt-zho* & \cmark &    \xmark &          TD &   G-Trans. &   Native &     Sparse & \cmark &   0.331 &      0.336 & 0.258 & 0.514 &  0.768 \\
                                         - &           coe22-bm25-t-mt-zho* & \cmark &    \xmark &           T &   G-Trans. &   Native &     Sparse & \cmark &   0.328 &      0.353 & 0.261 & 0.500 &  0.781 \\
jhu.mcnamee\cite{participants-jhu.mcnamee} &            jhumc.zhwords.td.rf & \xmark &    \xmark &          TD &    English &   Trans. &     Sparse & \cmark &   0.325 &      0.339 & 0.242 & 0.474 &  0.752 \\
jhu.mcnamee\cite{participants-jhu.mcnamee} &                jhumc.zh4.td.rf & \xmark &    \xmark &          TD &    English &   Trans. &     Sparse & \cmark &   0.323 &      0.334 & 0.256 & 0.475 &  0.726 \\
jhu.mcnamee\cite{participants-jhu.mcnamee} &             jhumc.zh5.td.ce.rf & \xmark &    \xmark &          TD &    English &   Trans. &     Sparse & \cmark &   0.319 &      0.364 & 0.258 & 0.443 &  0.626 \\
                                    h2oloo &     zh\_xdpr.mm.4rrf-mtQ.all.R & \xmark &    \xmark &          TD &      Other &   Native &      Dense & \cmark &   0.307 &      0.383 & 0.218 & 0.387 &  0.658 \\
                                    h2oloo &        \ml{zh\_dense-rrf.BM25} & \xmark &    \xmark &          TD &      Human &   Native &     Hybrid & \cmark &   0.302 &      0.308 & 0.210 & 0.442 &  0.706 \\
            KASYS\cite{participants-KASYS} &                  KASYS-run-zho & \xmark &    \xmark &           T &   G-Trans. &   Native &      Dense & \xmark &   0.286 &      0.311 & 0.166 & 0.341 &  0.525 \\
                                         - &                   \ml{zh\_2tr} & \xmark &    \xmark &          TD &      Human &     Both &     Sparse & \cmark &   0.284 &      0.279 & 0.203 & 0.463 &  0.702 \\
                                         - &                    \ml{zh\_2t} & \xmark &    \xmark &          TD &      Human &     Both &     Sparse & \xmark &   0.283 &      0.297 & 0.181 & 0.421 &  0.652 \\
                  F4\cite{participants-F4} &             F4-PyTerrierPL2-zh & \cmark &    \xmark &          TD &      Other &   Native &     Sparse & \xmark &   0.279 &      0.277 & 0.185 & 0.368 &  0.626 \\
                                         - &                   \ml{zh\_qtr} & \xmark &    \xmark &          TD &      Human &   Native &     Sparse & \cmark &   0.273 &      0.283 & 0.189 & 0.359 &  0.533 \\
                                         - &           coe22-bm25-d-mt-zho* & \cmark &    \xmark &           D &   G-Trans. &   Native &     Sparse & \cmark &   0.270 &      0.273 & 0.207 & 0.434 &  0.695 \\
                                         - &                    \ml{zh\_qt} & \xmark &    \xmark &          TD &      Human &   Native &     Sparse & \xmark &   0.257 &      0.285 & 0.173 & 0.335 &  0.494 \\
                                         - &                        zh\_dtr & \xmark &    \xmark &          TD &    English &   Trans. &     Sparse & \cmark &   0.252 &      0.238 & 0.180 & 0.370 &  0.579 \\
                                         - &                         zh\_dt & \xmark &    \xmark &          TD &    English &   Trans. &     Sparse & \xmark &   0.223 &      0.219 & 0.145 & 0.324 &  0.530 \\
              umcp\cite{participants-umcp} &                  umcp\_hmm\_zh & \xmark &    \xmark &           T &    English &   Native &     Sparse & \xmark &   0.220 &      0.217 & 0.147 & 0.343 &  0.632 \\
          IDACCS\cite{participants-IDACCS} &        IDACCS-run1\_rrank\_zho & \xmark &    \cmark &          TD &    English &   Native &      Dense & \xmark &   0.201 &      0.226 & 0.138 & 0.368 &  0.781 \\
                                         - &                      zh\_4rrf2 & \xmark &    \xmark &          TD &      Other &   Native &     Sparse & \cmark &   0.198 &      0.208 & 0.106 & 0.241 &  0.454 \\
                                         - &    IDACCS-baseline\_rrank\_zho & \xmark &    \cmark &          TD &    English &   Native &      Dense & \xmark &   0.192 &      0.203 & 0.131 & 0.389 &  0.781 \\
                                         - &                    zh\_4rrfprf & \xmark &    \xmark &          TD &      Other &   Native &     Sparse & \cmark &   0.190 &      0.203 & 0.107 & 0.245 &  0.456 \\
          IDACCS\cite{participants-IDACCS} &        IDACCS-run2\_rrank\_zho & \xmark &    \cmark &          TD &    English &   Native &      Dense & \xmark &   0.189 &      0.215 & 0.132 & 0.389 &  0.781 \\
                                         - &                       zh\_4rrf & \xmark &    \xmark &          TD &      Other &   Native &     Sparse & \xmark &   0.188 &      0.207 & 0.101 & 0.231 &  0.423 \\
          IDACCS\cite{participants-IDACCS} &               IDACCS-run2\_zho & \xmark &    \xmark &          TD &    English &   Native &      Dense & \xmark &   0.132 &      0.148 & 0.060 & 0.169 &  0.424 \\
          IDACCS\cite{participants-IDACCS} &               IDACCS-run1\_zho & \xmark &    \xmark &          TD &    English &   Native &      Dense & \xmark &   0.131 &      0.149 & 0.064 & 0.194 &  0.444 \\
                                         - &           IDACCS-baseline\_zho & \xmark &    \xmark &          TD &    English &   Native &      Dense & \xmark &   0.127 &      0.147 & 0.056 & 0.177 &  0.428 \\
                                    h2oloo &       zh\_xdpr.xorHn-mm.EN.d.R & \xmark &    \xmark &           D &    English &   Native &      Dense & \cmark &   0.122 &      0.144 & 0.085 & 0.193 &  0.354 \\
                                         - &         coe22-mhq-zho\_colxtt* & \cmark &    \xmark &         TDN &      Other &   Native &      Dense & \xmark &   0.036 &      0.053 & 0.025 & 0.044 &  0.571 \\
                                      RIET &            \ml{RietRandomRun2} & \xmark &    \xmark &           T &      Human &   Native &      Other & \xmark &   0.021 &      0.036 & 0.032 & 0.074 &  0.781 \\
\bottomrule
\multicolumn{14}{l}{Note that the original name of \texttt{NM.unicamp} was \texttt{unicamp}.}
\\\\ %
\end{tabular}
}

\end{table*}

%% file: _tab-full-results-fas.tex
\begin{table*}[]

\caption{Persian run results. Monolingual runs, which use human translations of the queries, are marked as green. Run used as the first stage retrieval for the reranking task is marked bold. * indicates manual runs.}\label{tab:fas-full-results}
\centering
\resizebox{\linewidth}{!}{
\begin{tabular}{ll|ccccccc|cccccc}
\toprule
       Team & Run ID &  Man. & ReR. &  QF &  QL & DL & Model &    PRF &  nDCG & RBP &    MAP & R@100 & R@1k \\
\midrule
  NM.unicamp\cite{participants-NM.unicamp} &                   p2.fa.rerank & \xmark &    \cmark &          TD &      Other &   Native &      Dense & \xmark &   0.588 &      0.479 & 0.435 & 0.681 &  0.829 \\
  NM.unicamp\cite{participants-NM.unicamp} &                \ml{p3.fa.mono} & \xmark &    \xmark &          TD &      Human &   Native &     Hybrid & \xmark &   0.562 &      0.462 & 0.422 & 0.706 &  0.833 \\
                NLE\cite{participants-NLE} &             NLE\_fa\_adhoc\_rr & \xmark &    \xmark &          TD &      Other &     Both &     Hybrid & \xmark &   0.551 &      0.455 & 0.406 & 0.677 &  0.926 \\
  NM.unicamp\cite{participants-NM.unicamp} &                      p4.fa.hoc & \xmark &    \xmark &          TD &      Other &   Native &     Hybrid & \cmark &   0.545 &      0.464 & 0.404 & 0.633 &  0.782 \\
  NM.unicamp\cite{participants-NM.unicamp} &                      p1.fa.hoc & \xmark &    \xmark &          TD &      Other &   Native &     Hybrid & \xmark &   0.536 &      0.449 & 0.397 & 0.666 &  0.807 \\
                NLE\cite{participants-NLE} &                 NLE\_fa\_adhoc & \xmark &    \xmark &          TD &   G-Trans. &   Native &     Hybrid & \xmark &   0.525 &      0.420 & 0.393 & 0.707 &  0.920 \\
                NLE\cite{participants-NLE} &             \ml{NLE\_fa\_mono} & \xmark &    \xmark &          TD &      Human &   Native &     Hybrid & \xmark &   0.523 &      0.443 & 0.400 & 0.698 &  0.913 \\
   CFDA\_CLIP\cite{participants-CFDA-CLIP} &                 CFDA\_CLIP\_dq & \xmark &    \cmark &          TD &    English &   Native &      Dense & \xmark &   0.508 &      0.436 & 0.364 & 0.607 &  0.758 \\
   CFDA\_CLIP\cite{participants-CFDA-CLIP} &             CFDA\_CLIP\_fas\_L & \xmark &    \xmark &          TD &    English &   Native &     Hybrid & \xmark &   0.488 &      0.420 & 0.343 & 0.589 &  0.758 \\
                NLE\cite{participants-NLE} &         \ml{NLE\_fa\_mono\_rr} & \xmark &    \xmark &          TD &      Human &   Native &     Hybrid & \xmark &   0.474 &      0.406 & 0.341 & 0.652 &  0.913 \\
   CFDA\_CLIP\cite{participants-CFDA-CLIP} &           CFDA\_CLIP\_fas\_clf & \xmark &    \xmark &          TD &    English &   Native &     Hybrid & \xmark &   0.468 &      0.405 & 0.330 & 0.590 &  0.758 \\
    huaweimtl\cite{participants-huaweimtl} &    \ml{huaweimtl-fa-m-hybrid1} & \xmark &    \xmark &          TD &      Human &    Other &     Hybrid & \xmark &   0.467 &      0.401 & 0.366 & 0.668 &  0.897 \\
                                         - &              splade\_farsi\_dt & \xmark &    \xmark &          TD &    English &   Trans. &   L-Sparse & \xmark &   0.455 &      0.372 & 0.287 & 0.588 &  0.833 \\
                                    h2oloo & \ml{fa\_dense-rrf.BM25.SPLADE} & \xmark &    \xmark &          TD &      Human &   Native &     Hybrid & \cmark &   0.446 &      0.393 & 0.327 & 0.685 &  0.917 \\
                                         - &              splade\_farsi\_mt & \xmark &    \xmark &          TD &   G-Trans. &   Native &   L-Sparse & \xmark &   0.444 &      0.369 & 0.314 & 0.621 &  0.835 \\
            KASYS\cite{participants-KASYS} &     KASYS\_onemodel-rerank-fas & \xmark &    \cmark &           T &    English &   Native &      Dense & \xmark &   0.415 &      0.372 & 0.285 & 0.546 &  0.829 \\
    huaweimtl\cite{participants-huaweimtl} &         huaweimtl-fa-c-hybrid3 & \xmark &    \xmark &          TD &      Other &   Trans. &     Hybrid & \xmark &   0.415 &      0.333 & 0.286 & 0.575 &  0.845 \\
                                         - &         \ml{splade\_farsi\_ht} & \xmark &    \xmark &          TD &      Human &   Native &   L-Sparse & \xmark &   0.415 &      0.360 & 0.291 & 0.599 &  0.825 \\
    huaweimtl\cite{participants-huaweimtl} &         huaweimtl-fa-c-hybrid2 & \xmark &    \xmark &         N/A &      Other &   Trans. &     Hybrid & \xmark &   0.411 &      0.339 & 0.280 & 0.578 &  0.845 \\
  hltcoe-jhu\cite{participants-hltcoe-jhu} &         coe22-tdq-fas\_colxtt* & \cmark &    \xmark &          TD &    English &   Native &      Dense & \xmark &   0.404 &      0.370 & 0.291 & 0.579 &  0.808 \\
  hltcoe-jhu\cite{participants-hltcoe-jhu} &          coe22-tq-fas\_colxtt* & \cmark &    \xmark &           T &    English &   Native &      Dense & \xmark &   0.395 &      0.339 & 0.273 & 0.568 &  0.773 \\
jhu.mcnamee\cite{participants-jhu.mcnamee} &                jhumc.fa4.td.rf & \xmark &    \xmark &          TD &    English &   Trans. &     Sparse & \cmark &   0.357 &      0.326 & 0.229 & 0.523 &  0.802 \\
                                         - & \textbf{coe22-bm25-td-dt-fas*} & \cmark &    \xmark &          TD &    English &   Trans. &     Sparse & \cmark &   0.355 &      0.315 & 0.253 & 0.517 &  0.829 \\
                                         - &                    \ml{fa\_qt} & \xmark &    \xmark &          TD &      Human &   Native &     Sparse & \xmark &   0.343 &      0.287 & 0.227 & 0.504 &  0.737 \\
                                         - &           coe22-bm25-t-dt-fas* & \cmark &    \xmark &           T &    English &   Trans. &     Sparse & \cmark &   0.341 &      0.309 & 0.232 & 0.539 &  0.797 \\
                                         - &                   \ml{fa\_qtr} & \xmark &    \xmark &          TD &      Human &   Native &     Sparse & \cmark &   0.341 &      0.297 & 0.236 & 0.521 &  0.809 \\
                                         - &            \ml{coe22-man-fas*} & \cmark &    \xmark &         TDN &      Human &   Native &     Sparse & \xmark &   0.337 &      0.312 & 0.231 & 0.476 &  0.795 \\
                                         - &           coe22-bm25-t-mt-fas* & \cmark &    \xmark &           T &   G-Trans. &   Native &     Sparse & \cmark &   0.334 &      0.295 & 0.221 & 0.486 &  0.786 \\
                                         - &                      fa\_3rrf2 & \xmark &    \xmark &          TD &      Other &   Native &     Sparse & \cmark &   0.332 &      0.282 & 0.219 & 0.495 &  0.767 \\
            KASYS\cite{participants-KASYS} &          KASYS\_one\_model-fas & \xmark &    \xmark &           T &    English &   Native &      Dense & \xmark &   0.330 &      0.310 & 0.200 & 0.434 &  0.591 \\
                                         - &          coe22-bm25-td-mt-fas* & \cmark &    \xmark &          TD &   G-Trans. &   Native &     Sparse & \cmark &   0.326 &      0.275 & 0.234 & 0.522 &  0.785 \\
                                         - &      \ml{coe22-bm25-t-ht-fas*} & \cmark &    \xmark &           T &      Human &   Native &     Sparse & \cmark &   0.326 &      0.290 & 0.222 & 0.498 &  0.788 \\
                                         - &                    fa\_3rrfprf & \xmark &    \xmark &          TD &      Other &   Native &     Sparse & \cmark &   0.326 &      0.278 & 0.215 & 0.494 &  0.782 \\
jhu.mcnamee\cite{participants-jhu.mcnamee} &                jhumc.fa5.td.rf & \xmark &    \xmark &          TD &    English &   Trans. &     Sparse & \cmark &   0.320 &      0.283 & 0.215 & 0.468 &  0.774 \\
                                         - &                       fa\_3rrf & \xmark &    \xmark &          TD &      Other &   Native &     Sparse & \xmark &   0.320 &      0.268 & 0.208 & 0.472 &  0.717 \\
                                         - &                    \ml{fa\_2t} & \xmark &    \xmark &          TD &      Human &     Both &     Sparse & \xmark &   0.317 &      0.263 & 0.231 & 0.485 &  0.751 \\
jhu.mcnamee\cite{participants-jhu.mcnamee} &             jhumc.fa5.td.ce.rf & \cmark &    \xmark &          TD &    English &   Trans. &     Sparse & \cmark &   0.316 &      0.306 & 0.209 & 0.381 &  0.602 \\
jhu.mcnamee\cite{participants-jhu.mcnamee} &            jhumc.fawords.td.rf & \xmark &    \xmark &          TD &    English &   Trans. &     Sparse & \cmark &   0.312 &      0.266 & 0.208 & 0.486 &  0.800 \\
          IDACCS\cite{participants-IDACCS} &         IDACCS-run1\_reranking & \xmark &    \cmark &          TD &    English &   Native &      Dense & \xmark &   0.311 &      0.281 & 0.189 & 0.533 &  0.829 \\
            KASYS\cite{participants-KASYS} &                      KASYS-run & \xmark &    \xmark &           T &   G-Trans. &   Native &      Dense & \xmark &   0.310 &      0.301 & 0.162 & 0.361 &  0.542 \\
                                         - &     \ml{coe22-bm25-td-ht-fas*} & \cmark &    \xmark &          TD &      Human &   Native &     Sparse & \cmark &   0.309 &      0.278 & 0.212 & 0.471 &  0.773 \\
                                    h2oloo &     fa\_xdpr.mm.2rrf-mtQ.all.R & \xmark &    \xmark &          TD &      Other &   Native &     Hybrid & \cmark &   0.309 &      0.296 & 0.203 & 0.465 &  0.693 \\
                                    h2oloo &         \ml{fa\_dense-rrf.prf} & \xmark &    \xmark &          TD &      Human &   Native &     Hybrid & \cmark &   0.308 &      0.292 & 0.205 & 0.491 &  0.735 \\
                                         - &           coe22-bm25-d-dt-fas* & \cmark &    \xmark &           D &    English &   Trans. &     Sparse & \cmark &   0.306 &      0.274 & 0.216 & 0.486 &  0.768 \\
    huaweimtl\cite{participants-huaweimtl} &       \ml{fa\_xdpr.ms.oht.d.R} & \xmark &    \xmark &           D &      Human &   Native &      Dense & \cmark &   0.299 &      0.286 & 0.192 & 0.431 &  0.661 \\
                                         - &                   \ml{fa\_2tr} & \xmark &    \xmark &          TD &      Human &     Both &     Sparse & \cmark &   0.291 &      0.248 & 0.202 & 0.487 &  0.770 \\
          IDACCS\cite{participants-IDACCS} &        IDACCS-run2\_rrank\_fas & \xmark &    \cmark &          TD &    English &   Native &      Dense & \xmark &   0.289 &      0.263 & 0.178 & 0.522 &  0.829 \\
                                         - &      \ml{coe22-bm25-d-ht-fas*} & \cmark &    \xmark &           D &      Human &   Native &     Sparse & \cmark &   0.281 &      0.224 & 0.197 & 0.444 &  0.742 \\
                                         - &           coe22-bm25-d-mt-fas* & \cmark &    \xmark &           D &   G-Trans. &   Native &     Sparse & \cmark &   0.268 &      0.217 & 0.183 & 0.445 &  0.691 \\
                                         - &     IDACCS-baseline\_raranking & \xmark &    \cmark &          TD &    English &   Native &      Dense & \xmark &   0.234 &      0.236 & 0.153 & 0.477 &  0.829 \\
              umcp\cite{participants-umcp} &                  umcp\_hmm\_fa & \xmark &    \xmark &           T &    English &   Native &     Sparse & \xmark &   0.206 &      0.176 & 0.121 & 0.417 &  0.650 \\
          IDACCS\cite{participants-IDACCS} &                    IDACCS-run1 & \xmark &    \xmark &          TD &    English &   Native &      Dense & \xmark &   0.188 &      0.173 & 0.107 & 0.337 &  0.557 \\
                  F4\cite{participants-F4} &                F4-PyTerrierPL2 & \cmark &    \xmark &          TD &      Other &   Native &     Sparse & \xmark &   0.181 &      0.166 & 0.111 & 0.330 &  0.577 \\
          IDACCS\cite{participants-IDACCS} &               IDACCS-run2\_fas & \xmark &    \xmark &          TD &    English &   Native &      Dense & \xmark &   0.165 &      0.156 & 0.094 & 0.306 &  0.529 \\
                                         - &                        fa\_dtr & \xmark &    \xmark &          TD &    English &   Trans. &     Sparse & \cmark &   0.163 &      0.134 & 0.122 & 0.328 &  0.485 \\
                                         - &                         fa\_dt & \xmark &    \xmark &          TD &    English &   Trans. &     Sparse & \xmark &   0.163 &      0.125 & 0.116 & 0.332 &  0.497 \\
                                         - &                IDACCS-baseline & \xmark &    \xmark &          TD &    English &   Native &      Dense & \xmark &   0.155 &      0.162 & 0.080 & 0.247 &  0.464 \\
                                    h2oloo &       fa\_xdpr.xorHn-mm.EN.d.R & \xmark &    \xmark &           D &    English &   Native &     Hybrid & \cmark &   0.140 &      0.136 & 0.087 & 0.284 &  0.544 \\
                                         - &         coe22-mhq-fas\_colxtt* & \cmark &    \xmark &         TDN &      Other &   Native &      Dense & \xmark &   0.032 &      0.041 & 0.022 & 0.064 &  0.705 \\
                                      RIET &             \ml{RietRandomRun} & \xmark &    \xmark &           T &      Human &   Native &      Other & \xmark &   0.032 &      0.042 & 0.027 & 0.059 &  0.829 \\
\bottomrule
\multicolumn{14}{l}{Note that the original name of \texttt{NM.unicamp} was \texttt{unicamp}.}
\end{tabular}
}

\end{table*}

%% file: _tab-full-results-rus.tex
\begin{table*}[]

\caption{Russian run results. Monolingual runs, which use human translations of the queries, are marked as green. Run used as the first stage retrieval for the reranking task is marked bold. * indicates manual runs.}\label{tab:rus-full-results}
\centering
\resizebox{\linewidth}{!}{
\begin{tabular}{ll|ccccccc|cccccc}
\toprule
       Team & Run ID &  Man. & ReR. &  QF &  QL & DL & Model &    PRF &  nDCG & RBP &    MAP & R@100 & R@1k \\
\midrule
  NM.unicamp\cite{participants-NM.unicamp} &                \ml{p3.ru.mono} & \xmark &    \xmark &          TD &      Human &   Native &     Hybrid & \xmark &   0.567 &      0.605 & 0.439 & 0.653 &  0.761 \\
                NLE\cite{participants-NLE} &             NLE\_ru\_adhoc\_rr & \xmark &    \xmark &          TD &      Other &     Both &     Hybrid & \xmark &   0.565 &      0.580 & 0.473 & 0.660 &  0.898 \\
  NM.unicamp\cite{participants-NM.unicamp} &                      p4.ru.hoc & \xmark &    \xmark &          TD &      Other &   Native &     Hybrid & \cmark &   0.563 &      0.594 & 0.434 & 0.643 &  0.804 \\
  NM.unicamp\cite{participants-NM.unicamp} &                      p1.ru.hoc & \xmark &    \xmark &          TD &      Other &   Native &     Hybrid & \xmark &   0.552 &      0.587 & 0.417 & 0.649 &  0.764 \\
                NLE\cite{participants-NLE} &         \ml{NLE\_ru\_mono\_rr} & \xmark &    \xmark &          TD &      Human &   Native &     Hybrid & \xmark &   0.550 &      0.546 & 0.429 & 0.610 &  0.885 \\
  NM.unicamp\cite{participants-NM.unicamp} &                   p2.ru.rerank & \xmark &    \cmark &          TD &      Other &   Native &      Dense & \xmark &   0.548 &      0.588 & 0.422 & 0.614 &  0.774 \\
   CFDA\_CLIP\cite{participants-CFDA-CLIP} &            CFDA\_CLIP\_rus\_dq & \xmark &    \xmark &          TD &    English &   Native &     Hybrid & \xmark &   0.513 &      0.542 & 0.386 & 0.526 &  0.701 \\
                                    h2oloo & \ml{ru\_dense-rrf.BM25.SPLADE} & \xmark &    \xmark &          TD &      Human &   Native &     Hybrid & \cmark &   0.510 &      0.523 & 0.385 & 0.602 &  0.894 \\
                NLE\cite{participants-NLE} &             \ml{NLE\_ru\_mono} & \xmark &    \xmark &          TD &      Human &   Native &     Hybrid & \xmark &   0.508 &      0.507 & 0.391 & 0.608 &  0.885 \\
   CFDA\_CLIP\cite{participants-CFDA-CLIP} &           CFDA\_CLIP\_rus\_clf & \xmark &    \xmark &          TD &    English &   Native &     Hybrid & \xmark &   0.507 &      0.518 & 0.383 & 0.536 &  0.701 \\
    huaweimtl\cite{participants-huaweimtl} &    \ml{huaweimtl-ru-m-hybrid1} & \xmark &    \xmark &          TD &      Human &   Trans. &     Hybrid & \xmark &   0.501 &      0.516 & 0.379 & 0.591 &  0.881 \\
    huaweimtl\cite{participants-huaweimtl} &         huaweimtl-ru-c-hybrid2 & \xmark &    \xmark &          TD &      Other &   Trans. &     Hybrid & \xmark &   0.496 &      0.515 & 0.377 & 0.578 &  0.879 \\
    huaweimtl\cite{participants-huaweimtl} &         huaweimtl-ru-c-hybrid3 & \xmark &    \xmark &          TD &      Other &   Trans. &     Hybrid & \xmark &   0.493 &      0.516 & 0.366 & 0.589 &  0.877 \\
                NLE\cite{participants-NLE} &                 NLE\_ru\_adhoc & \xmark &    \xmark &          TD &   G-Trans. &   Native &     Hybrid & \xmark &   0.490 &      0.512 & 0.384 & 0.591 &  0.877 \\
   CFDA\_CLIP\cite{participants-CFDA-CLIP} &             CFDA\_CLIP\_rus\_L & \xmark &    \xmark &          TD &    English &   Native &     Hybrid & \xmark &   0.469 &      0.482 & 0.349 & 0.517 &  0.701 \\
                                         - &            splade\_russian\_dt & \xmark &    \xmark &          TD &    English &   Trans. &   L-Sparse & \xmark &   0.467 &      0.468 & 0.350 & 0.544 &  0.823 \\
  hltcoe-jhu\cite{participants-hltcoe-jhu} &          coe22-tq-rus\_colxtt* & \cmark &    \xmark &           T &    English &   Native &      Dense & \xmark &   0.456 &      0.444 & 0.335 & 0.521 &  0.771 \\
  hltcoe-jhu\cite{participants-hltcoe-jhu} &         coe22-tdq-rus\_colxtt* & \cmark &    \xmark &          TD &    English &   Native &      Dense & \xmark &   0.451 &      0.475 & 0.328 & 0.526 &  0.784 \\
            KASYS\cite{participants-KASYS} &     KASYS\_onemodel-rerank-rus & \xmark &    \cmark &           T &    English &   Native &      Dense & \xmark &   0.450 &      0.472 & 0.321 & 0.490 &  0.774 \\
                                         - &       \ml{splade\_russian\_ht} & \xmark &    \xmark &          TD &      Human &   Native &   L-Sparse & \xmark &   0.432 &      0.453 & 0.302 & 0.511 &  0.778 \\
                                         - &            splade\_russian\_mt & \xmark &    \xmark &          TD &   G-Trans. &   Native &   L-Sparse & \xmark &   0.407 &      0.431 & 0.293 & 0.485 &  0.759 \\
                                    h2oloo &     ru\_xdpr.mm.2rrf-mtQ.all.R & \xmark &    \xmark &          TD &      Other &   Native &      Dense & \cmark &   0.404 &      0.419 & 0.276 & 0.452 &  0.691 \\
                                    h2oloo &         \ml{ru\_dense-rrf.prf} & \xmark &    \xmark &          TD &      Human &   Native &     Hybrid & \cmark &   0.404 &      0.415 & 0.259 & 0.424 &  0.696 \\
                                         - &                      ru\_2rrf2 & \xmark &    \xmark &          TD &      Other &   Native &     Sparse & \cmark &   0.371 &      0.429 & 0.263 & 0.457 &  0.788 \\
                                         - &                    ru\_2rrfprf & \xmark &    \xmark &          TD &      Other &   Native &     Sparse & \cmark &   0.370 &      0.419 & 0.264 & 0.465 &  0.804 \\
                                         - &                    \ml{ru\_qt} & \xmark &    \xmark &          TD &      Human &   Native &     Sparse & \xmark &   0.366 &      0.403 & 0.266 & 0.479 &  0.740 \\
            KASYS\cite{participants-KASYS} &          KASYS\_one\_model-rus & \xmark &    \xmark &           T &    English &   Native &      Dense & \xmark &   0.366 &      0.375 & 0.226 & 0.403 &  0.601 \\
                                         - &           coe22-bm25-t-mt-rus* & \cmark &    \xmark &           T &   G-Trans. &   Native &     Sparse & \cmark &   0.365 &      0.402 & 0.287 & 0.471 &  0.757 \\
                                         - &      \ml{coe22-bm25-t-ht-rus*} & \cmark &    \xmark &           T &      Human &   Native &     Sparse & \cmark &   0.363 &      0.398 & 0.288 & 0.454 &  0.759 \\
                                         - &     \ml{coe22-bm25-td-ht-rus*} & \cmark &    \xmark &          TD &      Human &   Native &     Sparse & \cmark &   0.363 &      0.404 & 0.279 & 0.487 &  0.766 \\
                                         - &                   \ml{ru\_qtr} & \xmark &    \xmark &          TD &      Human &   Native &     Sparse & \cmark &   0.362 &      0.416 & 0.272 & 0.483 &  0.774 \\
                                         - &          coe22-bm25-td-mt-rus* & \cmark &    \xmark &          TD &   G-Trans. &   Native &     Sparse & \cmark &   0.360 &      0.394 & 0.281 & 0.482 &  0.770 \\
    huaweimtl\cite{participants-huaweimtl} &       \ml{ru\_xdpr.ms.oht.d.R} & \xmark &    \xmark &           D &      Human &   Native &      Dense & \cmark &   0.354 &      0.349 & 0.237 & 0.390 &  0.591 \\
                                         - &            \ml{coe22-man-rus*} & \cmark &    \xmark &         TDN &      Human &   Native &     Sparse & \xmark &   0.347 &      0.398 & 0.254 & 0.413 &  0.680 \\
                                         - &                       ru\_2rrf & \xmark &    \xmark &          TD &      Other &   Native &     Sparse & \xmark &   0.346 &      0.403 & 0.237 & 0.478 &  0.744 \\
jhu.mcnamee\cite{participants-jhu.mcnamee} &             jhumc.ru5.td.ce.rf & \xmark &    \xmark &          TD &    English &   Trans. &     Sparse & \cmark &   0.342 &      0.349 & 0.225 & 0.389 &  0.565 \\
                                         - &           coe22-bm25-t-dt-rus* & \cmark &    \xmark &           T &    English &   Trans. &     Sparse & \cmark &   0.327 &      0.360 & 0.238 & 0.455 &  0.771 \\
                                         - &           coe22-bm25-d-dt-rus* & \cmark &    \xmark &           D &    English &   Trans. &     Sparse & \cmark &   0.327 &      0.360 & 0.238 & 0.455 &  0.771 \\
          IDACCS\cite{participants-IDACCS} &        IDACCS-run2\_rrank\_rus & \xmark &    \cmark &          TD &    English &   Native &      Dense & \xmark &   0.305 &      0.331 & 0.195 & 0.422 &  0.774 \\
                                         - &                    \ml{ru\_2t} & \xmark &    \xmark &          TD &      Human &     Both &     Sparse & \xmark &   0.305 &      0.306 & 0.216 & 0.425 &  0.699 \\
          IDACCS\cite{participants-IDACCS} &        IDACCS-run1\_rrank\_rus & \xmark &    \cmark &          TD &    English &   Native &      Dense & \xmark &   0.303 &      0.337 & 0.203 & 0.425 &  0.774 \\
                  F4\cite{participants-F4} &             F4-PyTerrierPL2-ru & \cmark &    \xmark &          TD &      Other &   Native &     Sparse & \xmark &   0.298 &      0.311 & 0.213 & 0.379 &  0.664 \\
                                         - & \textbf{coe22-bm25-td-dt-rus*} & \cmark &    \xmark &          TD &    English &   Trans. &     Sparse & \cmark &   0.292 &      0.324 & 0.216 & 0.425 &  0.774 \\
                                         - &                   \ml{ru\_2tr} & \xmark &    \xmark &          TD &      Human &     Both &     Sparse & \cmark &   0.290 &      0.311 & 0.216 & 0.442 &  0.733 \\
jhu.mcnamee\cite{participants-jhu.mcnamee} &            jhumc.ruwords.td.rf & \xmark &    \xmark &          TD &    English &   Trans. &     Sparse & \cmark &   0.283 &      0.314 & 0.202 & 0.435 &  0.657 \\
                                         - &           coe22-bm25-d-mt-rus* & \cmark &    \xmark &           D &   G-Trans. &   Native &     Sparse & \cmark &   0.282 &      0.301 & 0.204 & 0.383 &  0.673 \\
                                         - &    IDACCS-baseline\_rrank\_rus & \xmark &    \cmark &          TD &    English &   Native &      Dense & \xmark &   0.276 &      0.290 & 0.174 & 0.409 &  0.774 \\
jhu.mcnamee\cite{participants-jhu.mcnamee} &                jhumc.ru4.td.rf & \xmark &    \xmark &          TD &    English &   Trans. &     Sparse & \cmark &   0.276 &      0.303 & 0.198 & 0.398 &  0.649 \\
                                         - &      \ml{coe22-bm25-d-ht-rus*} & \cmark &    \xmark &           D &      Human &   Native &     Sparse & \cmark &   0.273 &      0.307 & 0.196 & 0.363 &  0.701 \\
jhu.mcnamee\cite{participants-jhu.mcnamee} &                jhumc.ru5.td.rf & \xmark &    \xmark &          TD &    English &   Trans. &     Sparse & \cmark &   0.263 &      0.284 & 0.199 & 0.411 &  0.704 \\
            KASYS\cite{participants-KASYS} &                  KASYS-run-rus & \xmark &    \xmark &           T &   G-Trans. &   Native &      Dense & \xmark &   0.257 &      0.279 & 0.151 & 0.288 &  0.511 \\
              umcp\cite{participants-umcp} &                  umcp\_hmm\_ru & \xmark &    \xmark &           T &    English &   Native &     Sparse & \xmark &   0.253 &      0.275 & 0.181 & 0.348 &  0.597 \\
                                    h2oloo &       ru\_xdpr.xorHn-mm.EN.d.R & \xmark &    \xmark &           D &    English &   Native &      Dense & \xmark &   0.193 &      0.208 & 0.110 & 0.232 &  0.471 \\
          IDACCS\cite{participants-IDACCS} &               IDACCS-run1\_rus & \xmark &    \xmark &          TD &    English &   Native &      Dense & \xmark &   0.184 &      0.180 & 0.093 & 0.235 &  0.498 \\
          IDACCS\cite{participants-IDACCS} &               IDACCS-run2\_rus & \xmark &    \xmark &          TD &    English &   Native &      Dense & \xmark &   0.172 &      0.174 & 0.082 & 0.225 &  0.487 \\
                                         - &           IDACCS-baseline\_rus & \xmark &    \xmark &          TD &    English &   Native &      Dense & \xmark &   0.154 &      0.170 & 0.075 & 0.185 &  0.468 \\
                                         - &                         ru\_dt & \xmark &    \xmark &          TD &    English &   Trans. &     Sparse & \xmark &   0.126 &      0.125 & 0.076 & 0.234 &  0.417 \\
                                         - &                        ru\_dtr & \xmark &    \xmark &          TD &    English &   Trans. &     Sparse & \cmark &   0.115 &      0.124 & 0.085 & 0.197 &  0.429 \\
                                         - &         coe22-mhq-rus\_colxtt* & \cmark &    \xmark &         TDN &      Other &   Native &      Dense & \xmark &   0.049 &      0.068 & 0.036 & 0.066 &  0.663 \\
                                      RIET &            \ml{RietRandomRun3} & \xmark &    \xmark &           T &      Human &   Native &      Other & \xmark &   0.035 &      0.058 & 0.038 & 0.093 &  0.774 \\
\bottomrule
\multicolumn{14}{l}{Note that the original name of \texttt{NM.unicamp} was \texttt{unicamp}.}
\end{tabular}
}
    
\end{table*}

%% file: _fig_topic_box.tex
\begin{figure*}
    \centering
    \includegraphics[width=0.9\linewidth]{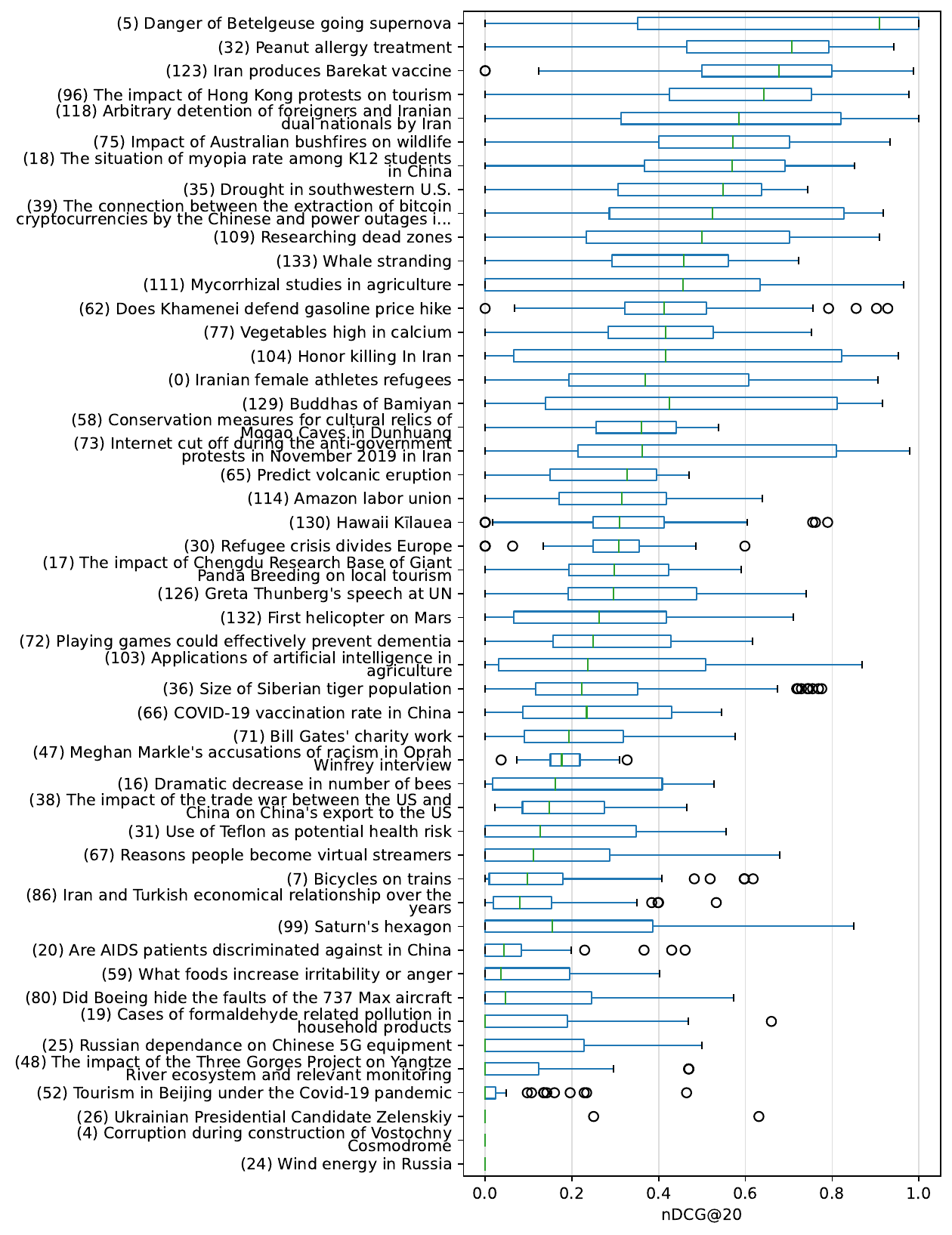}
    \caption{Boxplots of nDCG@20 on all Chinese runs.}
    \label{fig:topic-box-zho}
\end{figure*}

\begin{figure*}
    \centering
    \includegraphics[width=0.9\linewidth]{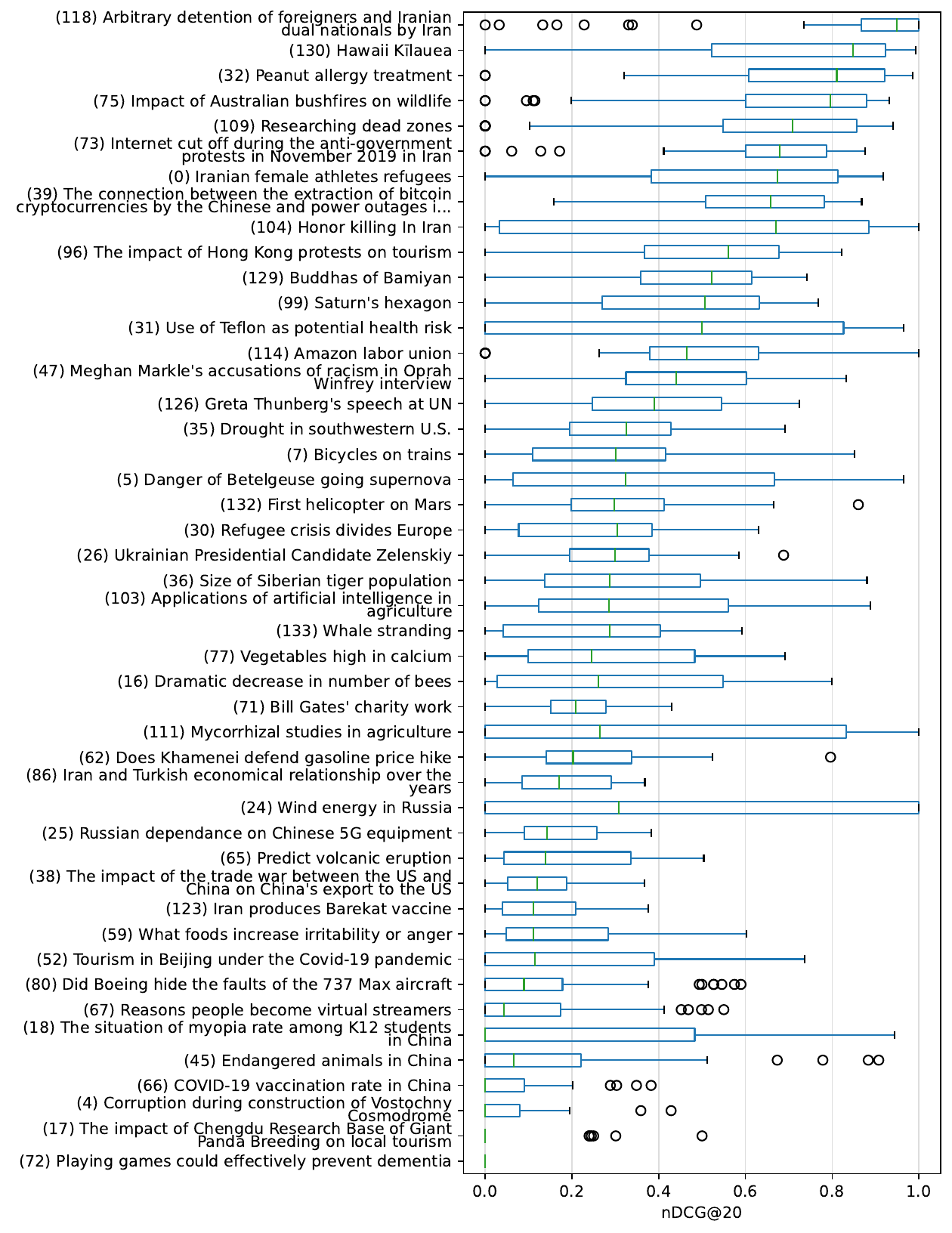}
    \caption{Boxplots of nDCG@20 on all Persian runs.}
    \label{fig:topic-box-fas}
\end{figure*}

\begin{figure*}
    \centering
    \includegraphics[width=0.9\linewidth]{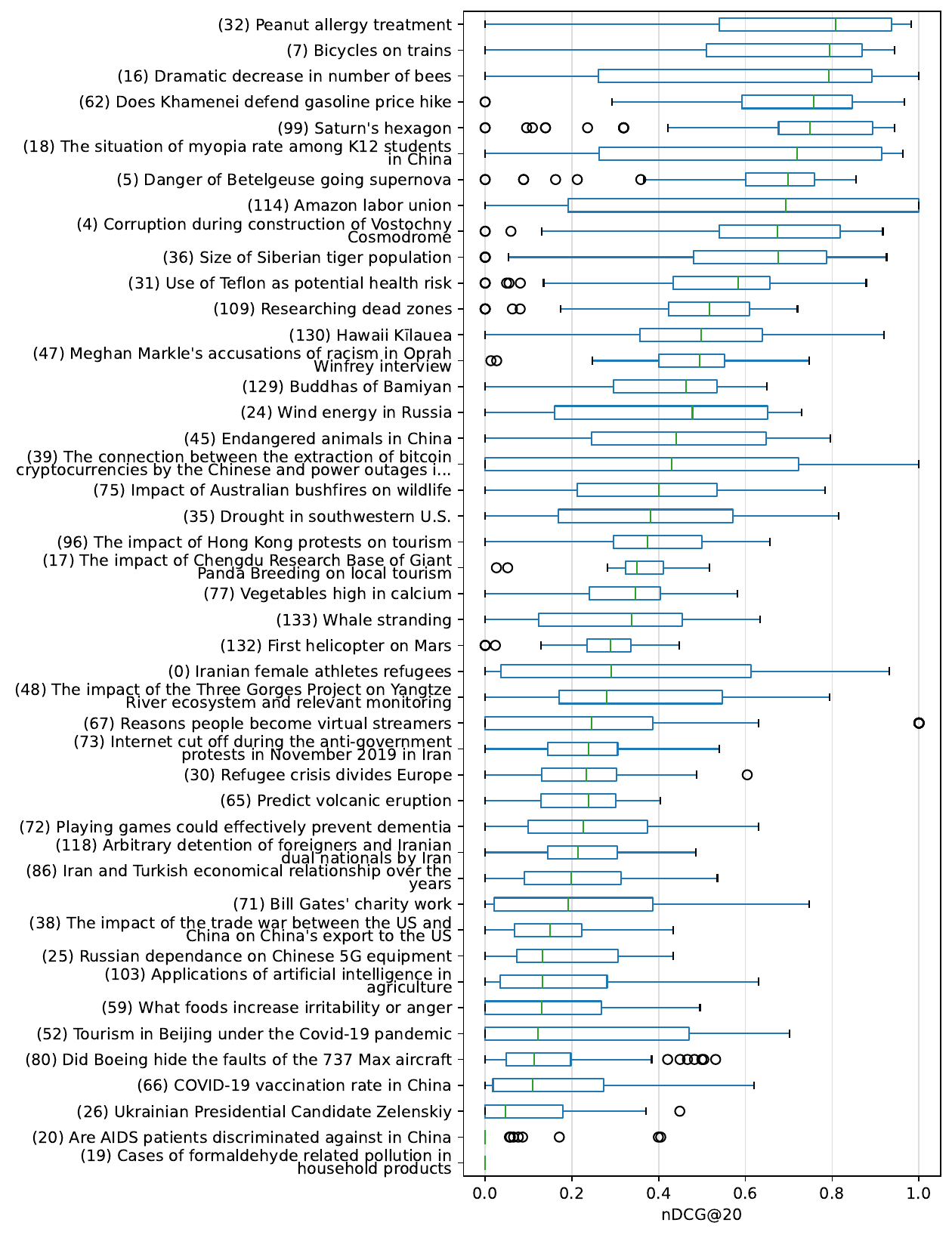}
    \caption{Boxplots of nDCG@20 on all Russian runs.}
    \label{fig:topic-box-rus}
\end{figure*}

%% file: _fig_overlaps.tex
\begin{figure*}
    \centering
    \includegraphics[width=0.85\linewidth]{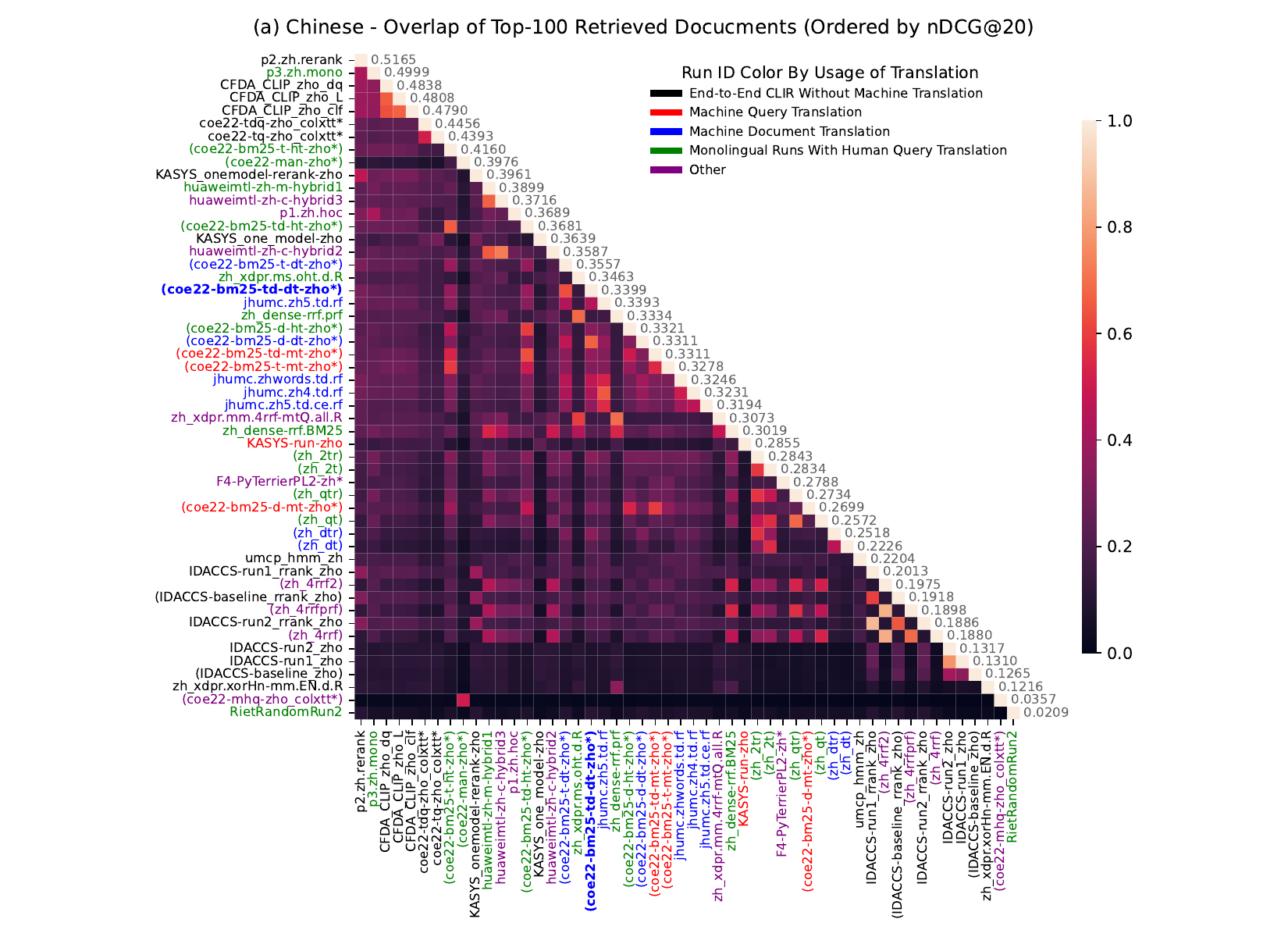}
    \includegraphics[width=0.85\linewidth]{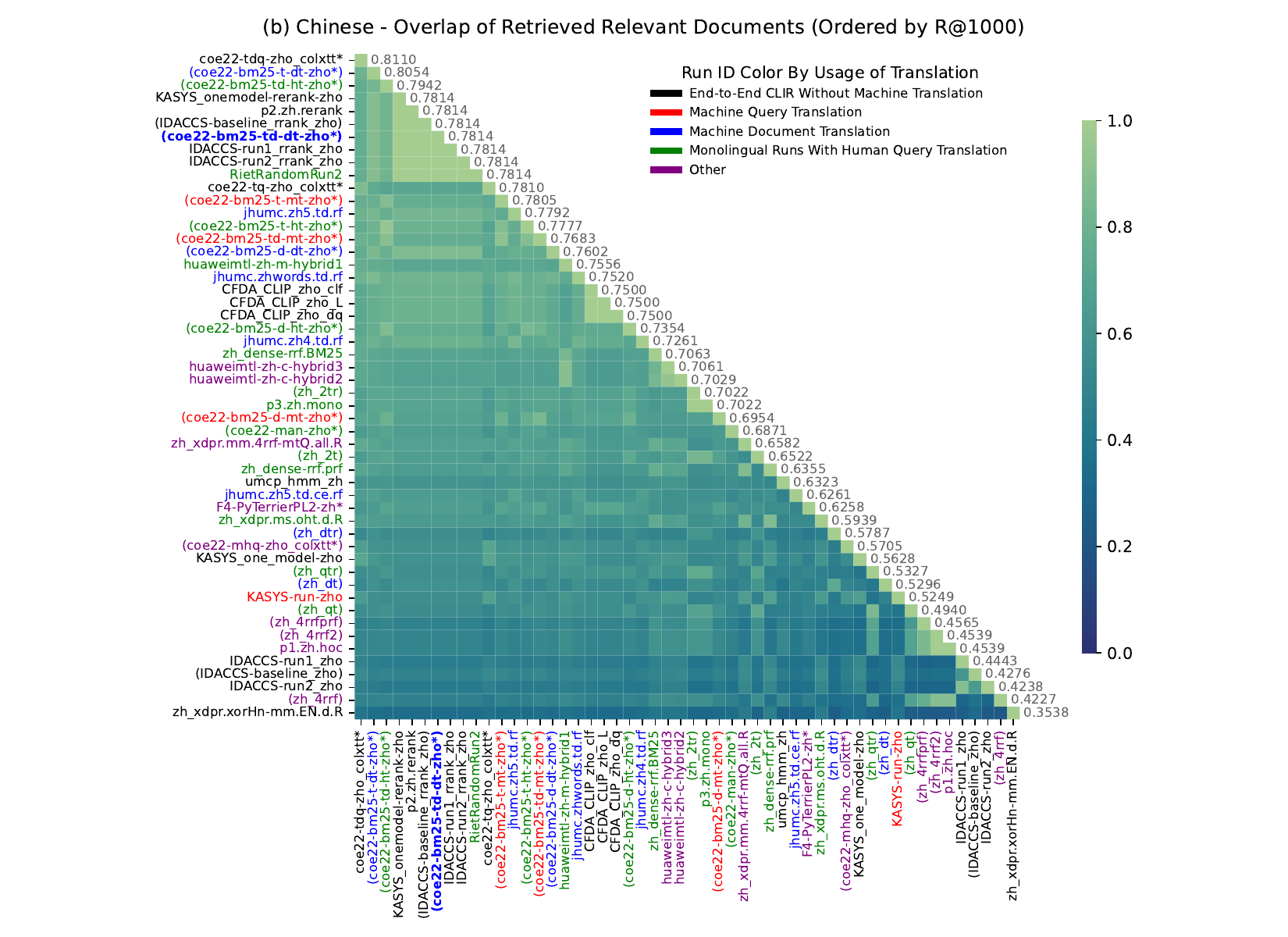}
    \caption{Overlap of documents retrieved by systems that participated in Chinese. Run used as the first stage retrieval for the reranking task is marked bold. * indicates manual runs. }
    \label{fig:zho-overlap}
\end{figure*}

\begin{figure*}
    \centering
    \includegraphics[width=0.85\linewidth]{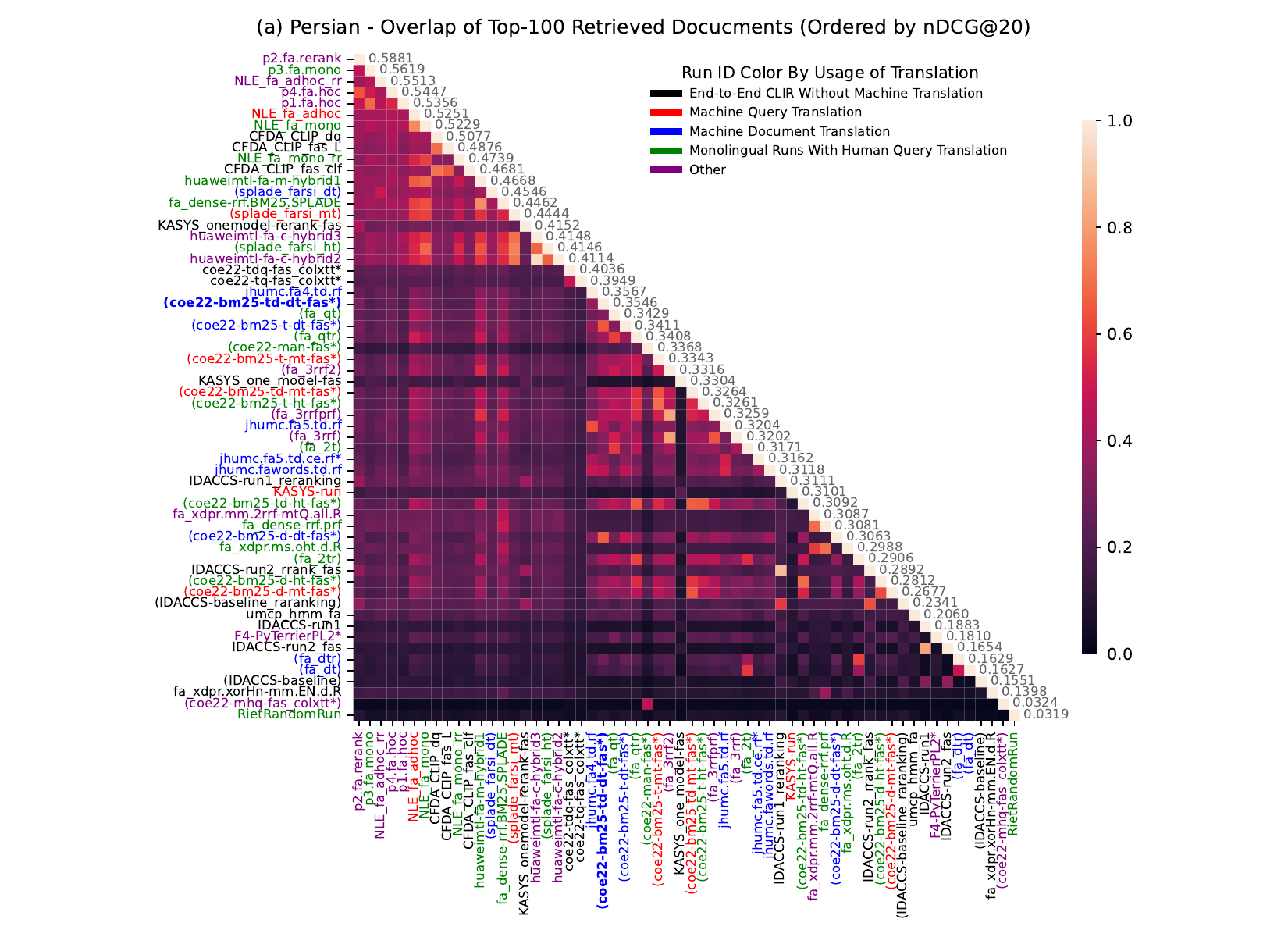}
    \includegraphics[width=0.85\linewidth]{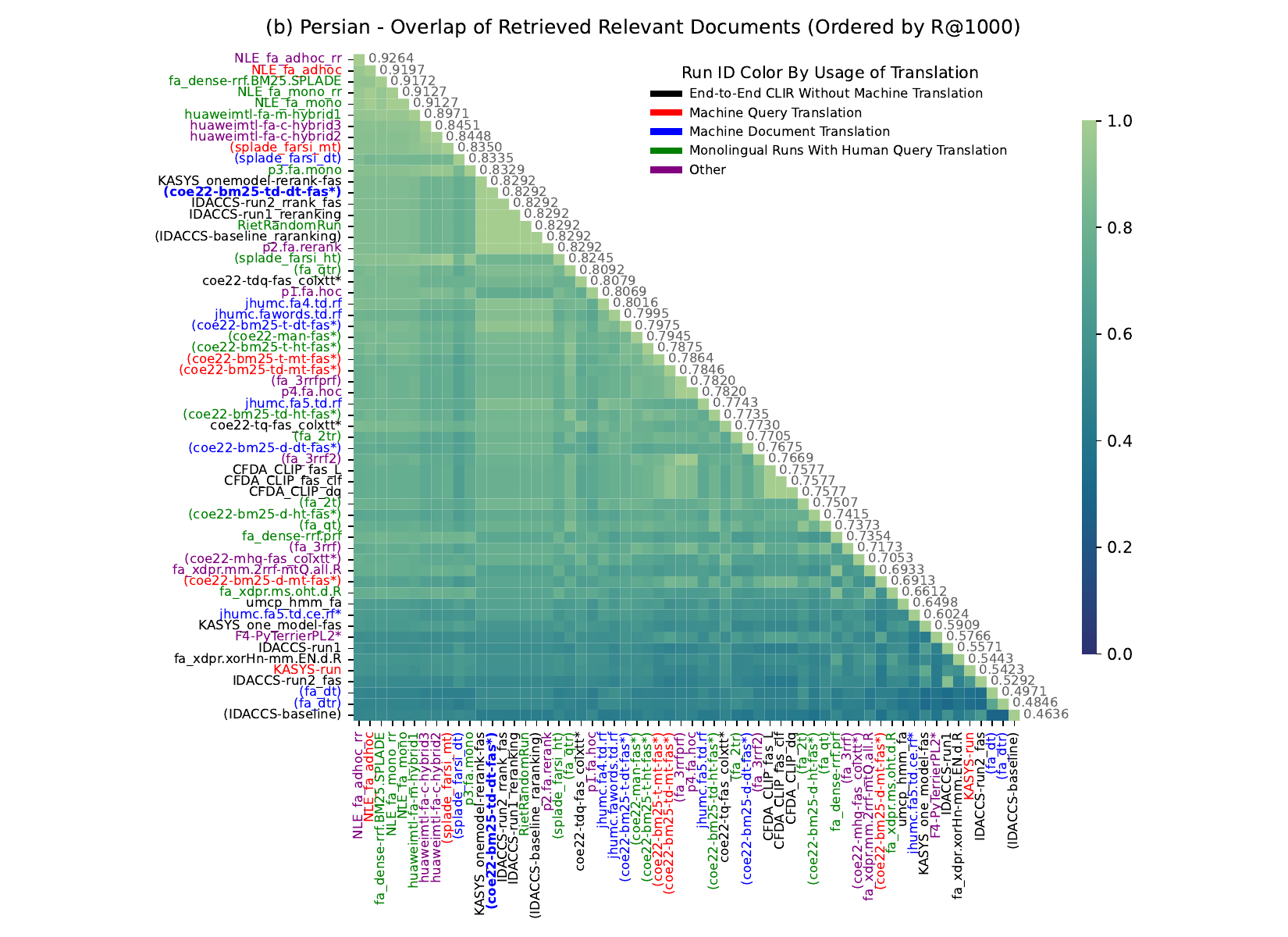}
    \caption{Overlap of documents retrieved by systems that participated in Persian. Run used as the first stage retrieval for the reranking task is marked bold. * indicates manual runs.}
    \label{fig:fas-overlap}
\end{figure*}

\begin{figure*}
    \centering
    \includegraphics[width=0.85\linewidth]{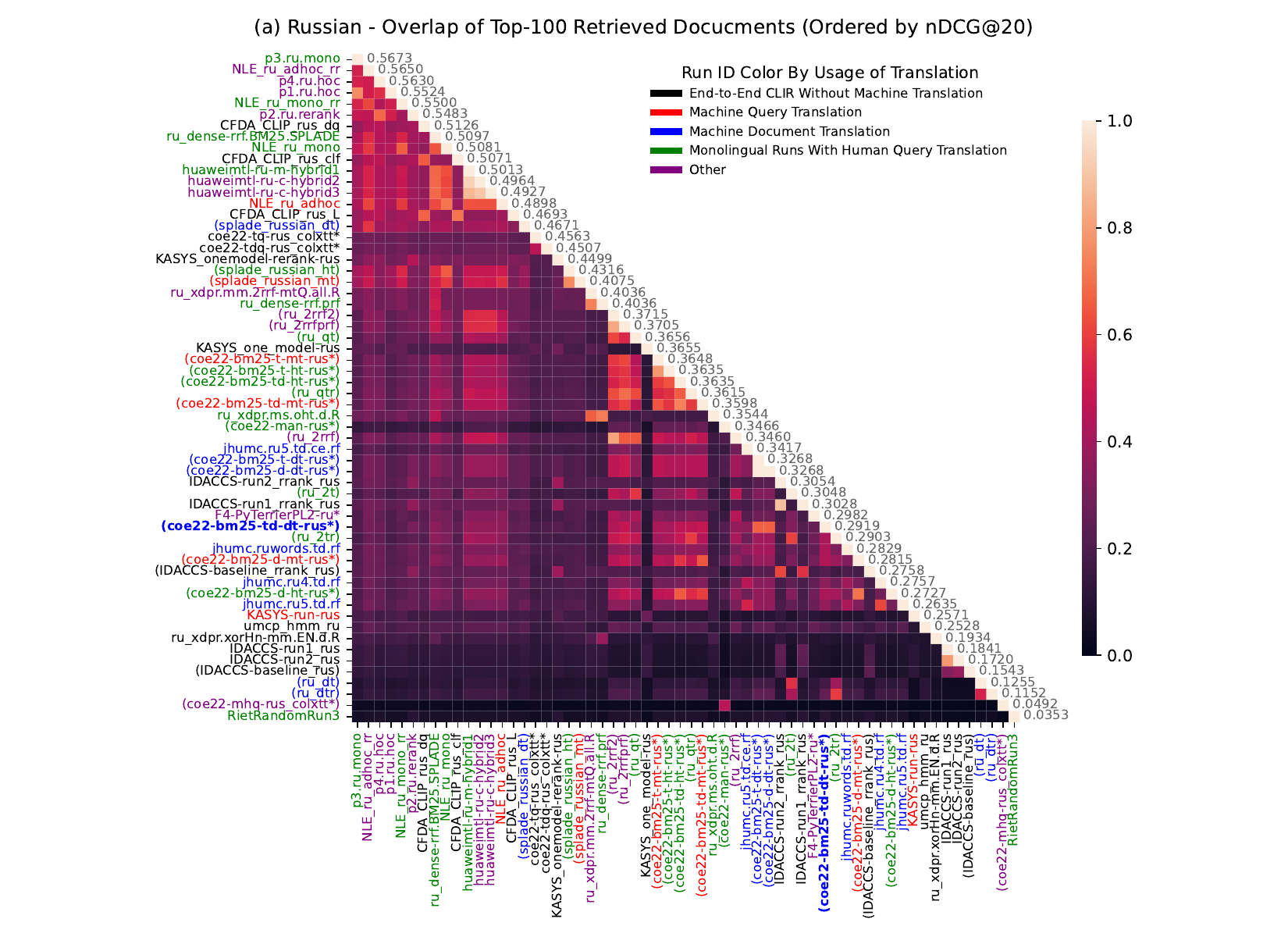}
    \includegraphics[width=0.85\linewidth]{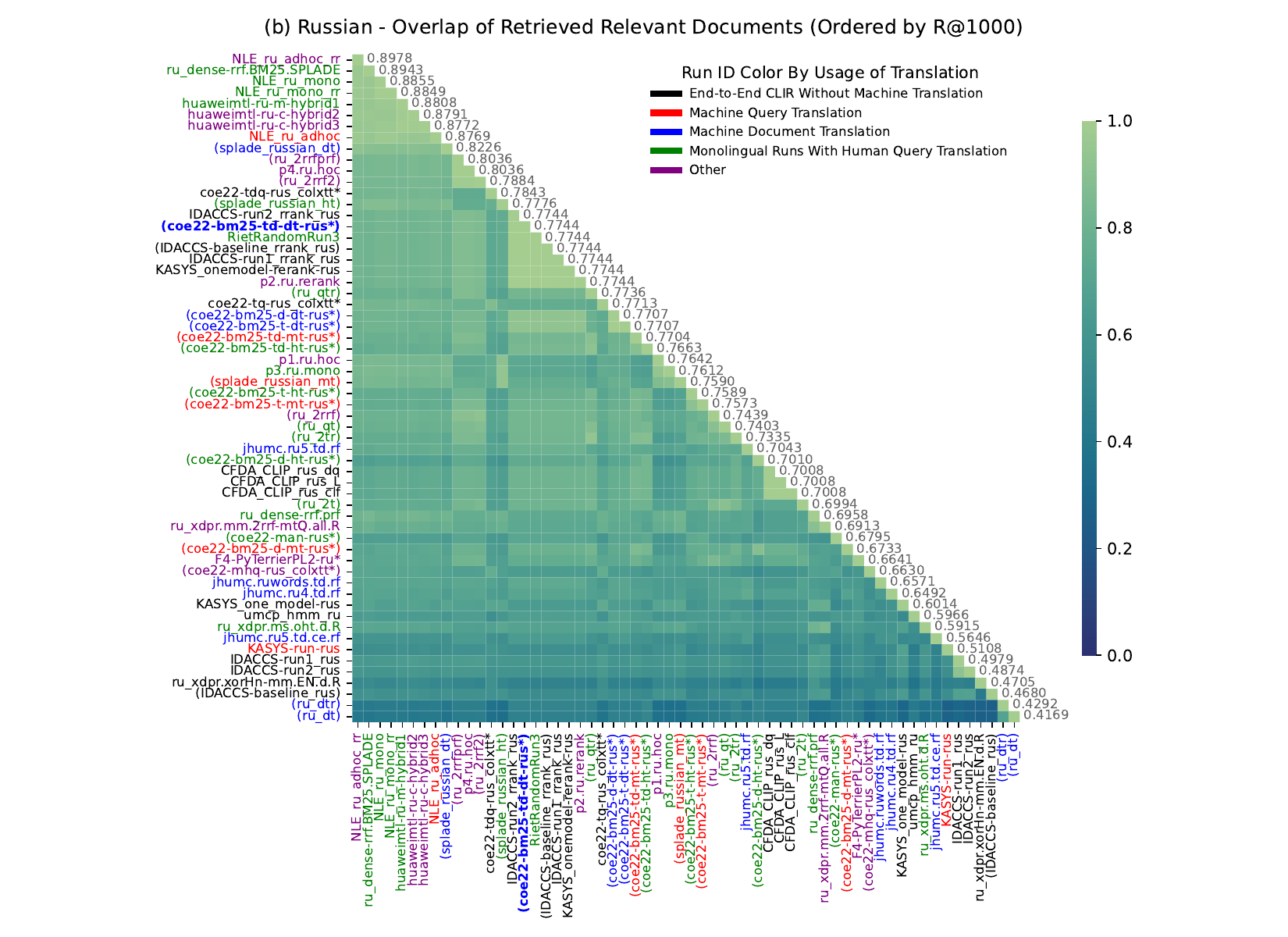}
    \caption{Overlap of documents retrieved by systems that participated in Russian. Run used as the first stage retrieval for the reranking task is marked bold. * indicates manual runs.}
    \label{fig:rus-overlap}
\end{figure*}